\documentclass[11pt]{article}
\usepackage[ansinew]{inputenc}
\usepackage{amssymb,amsfonts}
\usepackage{amsthm}
\usepackage{graphicx}

\def\ni{\noindent}
\def\nn{\nonumber}

\def \bc {\begin{center}}
\def \ec {\end{center}}
\def \bi {\begin{itemize}}
\def \ei {\end{itemize}}

\def \ba {\begin{array}}
\def \ea {\end{array}}

\def \bea {\begin{eqnarray}}
\def \eea {\end{eqnarray}}

\def \be {\begin{equation}}
\def \ee {\end{equation}}

\def\um {\frac{1}{2}}

\def\X {{\cal X}}
\def\pim{\frac{\pi}{2}}

\newtheorem{thm}{Theorem}[section]

\newtheorem{lem}[thm]{Lemma}
\newtheorem{prop}[thm]{Proposition}
\newtheorem{defn}[thm]{Definition}
\theoremstyle{remark}

\newcommand{\parcial}[1]{ \frac{\partial}{\partial #1} }

\begin{document}
\begin{center}
{\LARGE {\bf Wavelet Transform on the Circle and the Real Line:
\\ \vskip 3mm a Unified Group-Theoretical Treatment \footnote{Work
partially supported by the MCYT and Fundación Séneca under projects BFM
2002-00778 and PB/9/FS/02}}}
\end{center}
\bigskip
\bigskip

\centerline{{\sc M. Calixto$^1$ and J. Guerrero$^2$}}

\bigskip

\bc {\it $^1$Departamento de Matemática Aplicada y Estad\'\i stica,
Universidad Politécnica de Cartagena, Paseo Alfonso XIII 56, 30203
Cartagena, Spain}\\ and \\ {\it $^2$Departamento de Matemática Aplicada,
Universidad de Murcia, Facultad de Informática, Campus de Espinardo,
30100 Murcia, Spain}
%{\it Instituto de Astrof\'\i sica de
%Andaluc\'\i a, Apartado Postal 3004, 18080 Granada, Spain}
%\\ %and \\
%{\it Instituto Carlos I de F\'\i sica Teórica y Computacional, Facultad
%de Ciencias, Universidad de Granada, Campus de Fuentenueva, %Granada 18002, Spain.}

\bigskip

 Manuel.Calixto@upct.es \hskip 1cm juguerre@um.es\ec

\bigskip

\bigskip
\begin{center}
{\bf Abstract}
\end{center}
\small

\begin{list}{}{\setlength{\leftmargin}{3pc}\setlength{\rightmargin}{3pc}}
\item We present a unified group-theoretical derivation of the Continuous
Wavelet Transform (CWT) on the circle $\mathbb S^1$ and the real line
$\mathbb{R}$, following the general formalism of Coherent States
(CS) associated to unitary square integrable (modulo a subgroup, possibly)
representations of the group $SL(2,\mathbb{R})$. A general procedure
for obtaining unitary representations of a group $G$ of affine transformations
on a space of signals $L^2(X,dx)$ is described, relating carrier spaces $X$ to
(first or higher-order) ``polarization subalgebras'' ${\cal P}_X$.
We also provide explicit admissibility and continuous frame
conditions for wavelets on $\mathbb S^1$ and discuss the Euclidean limit
in terms of group contraction.
\end{list}
\normalsize %\setlength{\baselineskip}{14pt}

%\noindent PACS:

%\noindent MSC:

%\section*{Falta por ver:}
%
%\begin{itemize}
%\item Condicion de continuous frame, segunda parte.
%\item Contraccion respecto al subgrupo parabolico (no conexo): $SIM(1)\times Z_2$
%(que pasa con los modos negativos?)
%\item Considerar todos los valores de la serie continua $1/2+is$
%\item Considerar la serie continua $C_q^{1/2}$.
%\end{itemize}
%
%
%\noindent {\bf Keywords:}

%\newpage
\section{Introduction}

It is not straightforward to define a proper dilation operator on the
circle, and therefore the wavelets on the circle cannot be attained by an
irreducible representation of the affine group in a straightforward way.
Holschneider obtained in Ref. \cite{Holschneider-circle} wavelets on the
circle from the standard ones $\gamma_{b,a}(x)=a^{-1}\gamma(\frac{x-b}{a})$ by
means of periodization: \be
\gamma_{\theta,a}(x)=\sum_{n\in\mathbb{Z}}\frac{1}{a}\gamma\left(\frac{x-\theta+2\pi
n}{a}\right),\; \theta\in \mathbb S^1, a\in \mathbb{R}^+,\ee for functions
$\gamma$ which decay sufficiently fast at infinity.

In Ref. \cite{Holschneider-sphere}, the same author proposed a CWT on the
sphere $\mathbb{S}^2$ satisfying some natural requirements, but with a number of
{\it ad hoc} assumptions (in particular, for dilations).
Later, in Ref. \cite{waveS2}, this proposal was derived, and the assumptions proved,
in a group-theoretical setting, by means of an appropriate unitary representation of
the Lorentz group $SO_0(3,1)$, using the coherent state machinery (see \cite{Gazeau} for a review).
This construction was extended afterwards to higher dimensional spheres
$\mathbb{S}^{n-1}$ in \cite{nsphere} by using the Lorentz group in $n+1$ dimensions
$SO_0(n,1)$. However, the singular case of $n=2$ (corresponding to the circle
$\mathbb{S}^1$) does not fit in with the general case.

Here we identify the group of affine transformations of $\mathbb S^1$ and
$\mathbb{R}$, namely $SL(2,\mathbb{R})$, and derive the CWT on both spaces
in a unified manner, based on the construction of general CS associated to
square integrable representations (modulo a subgroup \cite{Gazeau}) of $SL(2,\mathbb{R})$.

To that purpose, we outline a powerful technique (explained in Sec.
\ref{obtaining}) for obtaining representations of Lie groups, which is a
mixture of Mackey's induction procedure and Kostant-Kirillov coadjoint
orbits method. The advantage of following this general procedure is that
it naturally reproduces a number of \emph{ad hoc} assumptions found in the
literature (i.e.: concerning semi-invariance and multipliers, definition
of dilations, Euclidean limits, etc) to define a genuine CWT on a manifold
$X$. This group-approach to the CWT also unifies representations in different carrier spaces $X$,
associating them to different ``polarization subalgebras'' ${\cal P}_X$
(see later on Sec. \ref{obtaining} for a formal definition), and providing
unitary isomorphisms or ``polarization-changing operators'' (e.g.: Fourier
and Laplace Transforms) which relate them. We shall explain this technique
with the help of the affine group in one dimension. This simple example
will serve us as a preamble to tackle the more involved case of
$SL(2,\mathbb R)$, where affine and ``circular'' wavelets coexist as different
representation spaces associated to non-equivalent polarization
subalgebras.

The organization of the paper is as follows. In Section \ref{brief} we
briefly sketch the construction of coherent states associated to group
representations, following \cite{Gazeau}. In Section \ref{obtaining} we describe a general procedure
for obtaining representations of a group $G$ of affine transformations on
a space of signals $L^2(X,dx)$, by relating carrier spaces $X$ to
``polarization subalgebras'' ${\cal P}_X\subset{\cal G}$, the Lie algebra
of $G$. Inside this scheme, Fourier and Laplace transforms will appear as
particular cases of what we define in general by: ``polarization-changing
operators". As already said, we exemplify this abstract construction with
the simple case of the affine group in Section \ref{afin}. In Section \ref{sl2r} we derive
discrete and continuous series representations of
$SL(2,\mathbb R)$, relating them to wavelets on the real line and the
circle, respectively. Associated to two different polarization
subalgebras, ${\cal P}_{\mathbb R^+}$ and ${\cal P}_{\mathbb C^+}$, we
realize discrete series (affine wavelets) on the half line
$\mathbb R^+$ (Section \ref{sl2rRmas}) and on the right half complex plane $\mathbb C^+$
(Section \ref{sl2rCmas}) and provide the corresponding polarization-changing operator (which turns out to be the Laplace transform) in terms
of the generating function of the generalized Laguerre polynomials (the
kernel). A third (non-equivalent) polarization subalgebra ${\cal
P}_{\mathbb S^1}$ of
$sl(2,\mathbb R)$ leads to a realization of continuous series representations on the
circumference (Section \ref{contser}); here we derive the CWT on $\mathbb
S^1$, using the coherent state machinery explained in Sec. \ref{brief},
from an appropriate unitary representation of $SL(2,\mathbb R)$ obtained
by following the general procedure designed in Sec. \ref{obtaining}. In
particular, we provide admissibility conditions and prove that the
corresponding family of coherent states constitutes a continuous frame, following
the strategy of \cite{waveS2} for $\mathbb{S}^2$. We
also study the Euclidean limit and prove that the continuous series
representation of $SL(2,\mathbb R)$ on $\mathbb S^1$ contract to the usual
wavelet representation of the affine group on $\mathbb R$ in the limit
$R\to \infty$ (large radius).

\section{The CWT on a Manifold\label{brief}}

The usual CWT on the real line $X=\mathbb{R}$ is derived from the natural
unitary representation of the affine group $G=SIM(1)$ in the space of
finite energy signals $L^2(\mathbb{R}, dx)$. The same scheme applies to
the CWT on a general manifold $X$, subject to the transitive action, $x\to
gx, g\in G, x\in X$, of some group of transformations $G$ which contains
dilations. If the measure $dx$ is $G$-invariant (i.e. $d(gx)=dx$), then
the natural left action of $G$ on $L^2(X,dx)$ given by:
\be
[U^L(g)\psi](x)=\psi(g^{-1}x), \;\;g\in G, \psi\in L^2(X,dx),\label{repre}
\ee defines a unitary representation, that is: \be \langle
U^L(g)\psi|U^L(g)\phi\rangle=\langle\psi|\phi\rangle\equiv\int_X\bar{\psi}(x)\phi(x)dx.\ee
When $dx$ is not strictly invariant (i.e.
$d(gx)=\lambda(g,x)dx$), we have to introduce a \emph{multiplier}
\be
[U^L(g)\psi](x)=\lambda(g,x)^{1/2}\psi(g^{-1}x), \;\;g\in G, \psi\in
L^2(X,dx), \label{multiplier}\ee
in order to keep unitarity. The fact that
$U^L(g_2)U^L(g_1)=U^L(g_2g_1)$ (i.e. $U^L$ is a representation of $G$)
implies cohomology conditions for multipliers, that is:
\begin{equation}
\lambda(g_2g_1,x)=\lambda(g_2,x)\lambda(g_1,g_2^{-1}x).  \label{1-cocycle}
\end{equation}
In this case, multipliers are associated with the Radon-Nikodym derivative of the quasi-invariant measure.
However, we shall show how multipliers naturally emerge from our scheme in a
non-ad-hoc manner (see next Section).

Let us consider the space $L^2(G,d^Lg)$ of square-integrable complex
functions $\Psi$ on $G$, where $d^Lg=d^L(g'g),\,\forall g'\in G$, stands
for the left-invariant Haar measure, which defines the scalar product: \be
\left(\Psi|\Phi\right)=\int_G\bar{\Psi}(g)\Phi(g)d^Lg. \ee
A non-zero function $\gamma\in L^2(X,dx)$ is called \emph{admissible} (also ``wavelet
analyzing function") if
$\Gamma(g)\equiv \langle U^L(g)\gamma|\gamma\rangle\in L^2(G,d^Lg)$, that is, if
\be C_\gamma=\int_G\bar{\Gamma}(g)\Gamma(g)d^Lg=\int_G|\langle
U^L(g)\gamma|\gamma\rangle|^2d^Lg<\infty. \label{norm}\ee

Let us assume that the representation $U^L$ is \emph{irreducible}, and that there exists a function
$\gamma$ admissible, then a system of coherent states (CS) on $X$ associated to (or indexed by)
$G$ are defined as the functions in the orbit of $\gamma$ under $G$:
\be
\gamma_g=U^L(g)\gamma, \;\; g\in G. \ee

When the representation $U^L$ is
not square integrable, it is not possible to find admissible functions $\gamma$ since
$\Gamma(g)$ is not square integrable. We can still proceed by restricting ourselves to a
suitable homogeneous space $Q=G/H$, for some closed subgroup $H$. Then,
the non-zero function $\gamma$ is said to be admissible mod$(H,\sigma)$
(with
$\sigma:Q\to G$ a Borel section), and the representation $U^L$ square integrable
mod$(H,\sigma)$, if the following condition holds:
\be
\int_Q|\langle U^L(\sigma(q))\gamma|\psi\rangle|^2 d^Lq<\infty,\;\;\forall
\psi\in L^2(X,dx),\label{qsquare}\ee where $d^Lq$ is a measure on $Q$
``projected'' from the left-invariant measure $d^Lg$ on the whole $G$. The
coherent states indexed by $Q$ are defined as
$\gamma_{\sigma(q)}=U^L(\sigma(q))\gamma, q\in Q$, and they form an overcomplete
set in $L^2(X,dx)$.

The condition (\ref{qsquare}) could also be written as an expectation value
\be
0<\int_Q |\langle U^L(\sigma(q))\gamma|\psi\rangle|^2 d^Lq=\langle
\psi|A_\sigma |\psi\rangle <\infty ,\;\;\forall \psi\in
L^2(X,dx),\label{pbiop}\ee where
$A_\sigma=\int_Q|\gamma_{\sigma(q)}\rangle\langle \gamma_{\sigma(q)}|d^Lq$ is a positive, bounded,
invertible operator. If the operator $A_\sigma^{-1}$ is also bounded, then
the set $S_\sigma=\{\gamma_{\sigma(q)}, q\in Q\}$ is called a
\emph{frame}, and a \emph{tight frame} if $A_\sigma$ is a positive
multiple of the identity, $A_\sigma=\lambda \mathbb{I}, \lambda>0$.

To avoid domain problems in the following, let us assume that $\gamma$
generates a frame (i.e. that $A_\sigma^{-1}$ is bounded). The \emph{CS
map} or \emph{wavelet transform} is defined as the linear map
\be\begin{array}{cccc} T_\gamma: & L^2(X,dx) &\longrightarrow&
L^2(Q,d^Lq)\\
 & \psi & \longmapsto & \Psi_\gamma(q)=[T_\gamma\psi](q)=\frac{\langle
\gamma_{\sigma(q)}|\psi\rangle}{\sqrt{c_\gamma}},\end{array}
\label{cwt}\ee where
$c_\gamma$ is the squared norm of $\Gamma(q)\equiv \langle U^L(\sigma(q))\gamma|\gamma\rangle\in
L^2(Q,d^Lq)$.
%This normalization makes the wavelet transform an
%\emph{isometry}, that is, it fulfils the following identity: \be
%\langle\psi|\phi\rangle=\int_Q\overline{[T_\gamma\psi]}(q)
%[T_\gamma\phi](q)d^Lq\equiv
%\left(\Psi_\gamma|\Phi_\gamma\right)_Q .\ee
Its range $L^2_\gamma(Q,d^Lq)\equiv T_\gamma(L^2(X,dx))$ is complete with respect to the scalar product
$(\Phi|\Psi)_\gamma\equiv\left(\Phi|T_\gamma A_\sigma^{-1} T_\gamma^{-1}\Psi\right)_Q$
and $T_\gamma$ is unitary from $L^2(X,dx)$ onto $L^2_\gamma(Q,d^Lq)$.
Thus, the inverse map $T_\gamma^{-1}$ yields the \emph{reconstruction
formula}:
\be
\psi=T_\gamma^{-1}\Psi_\gamma=\int_Q\Psi_\gamma(q)A_\sigma^{-1}\gamma_{\sigma(q)}
d^Lq,\;\;\Psi_\gamma\in L^2_\gamma(Q,d^Lq),\ee which expands the signal
$\psi$ in terms of CS $A_\sigma^{-1}\gamma_{\sigma(q)}$ with wavelet
coefficients
$\Psi_\gamma(q)=[T_\gamma\psi](q)$. These formulas acquire a simpler form when $A_\sigma$ is a multiple
of the identity.

\section{Obtaining the representations of $G$ on $L^2(X,dx)$ \label{obtaining}}

%\section{Obtaining $L^2(X,dx)$ from $L^2(G,d^Lg)$}

%Aqui tenemos que discutir el caso en que $L^2(X,dx)$ se obtiene a partir
%de la representación regular $L^2(G,d^{L,R}g)$ como subproducto de una
%polarización izquierda o derecha (creo que esto es algo nuevo que nosotros
%aportamos). Para el caso de las wavelets sobre la circunferencia, me huelo
%que, siendo la serie continua, no va a ser cuadrado integrable y tendremos
%que usar una sección $\sigma$ como en la sección anterior, para quitarnos
%el parámetro $b\in R$.
%
%Bueno, en esta sección podemos hablar de los campos derechos e izquierdos,
%de las polas, etc, y aplicarlo al caso del grupo afín, dejando $SL(2,R)$
%para la siguiente sección.

In this section we shall explain the procedure to obtain the irreducible
representations of a group $G$ on a space $L^2(X,dx)$, where
$X=G/H$ is an homogeneous space under $G$, from the left action of $G$ on
complex (Borel) functions $\Psi$ over $G$, ${\cal F}(G)$:
\be
[U^L(g)\Psi](g')=\Psi(g^{-1}g')\,,\qquad \Psi\in {\cal
F}(G)\label{regular} \ee

This procedure is a mixture of Mackey's induction technique and
Kostant-Kirillov coadjoint orbit method, with some new ingredients such as
higher-order polarizations. It is known as Group Approach to Quantization
(GAQ) \cite{JMP23,CMM}, although in this paper we shall use
non-horizontal polarizations instead of pseudo-extensions
\cite{JMP-pseudo}.

The idea is to consider the representation induced by a certain one-dimensional representation
$D_\alpha$ of a subgroup
$P$ of $G$, restricting the representation (\ref{regular}) to the subspace
${\cal H}_\alpha\subset {\cal F}(G)$ of (Borel) functions on $G$ satisfying:
\be
\Psi(gh)=D_\alpha(h)\Psi(g)\,\,,\forall g\in G\,,\forall h\in P \label{inducida}
\ee

We shall
%restrict ourselves to one-dimensional representations $D_\alpha$ of $P$, and
choose $P$ appropriately to obtain an irreducible representation of $G$ (see
below about irreducibility).

For simplicity we shall
consider connected and simply connected Lie groups. In the case the
group is not simply connected, we study the representations of the universal covering group and then
determine which representations of the universal covering group are also representations of the original group.

We shall also suppose that $P$ is connected and simply connected. Under
this conditions, we can work at the infinitesimal level and use left
invariant vector fields
%(which generate the finite right action $[U^R(g')\Psi](g)=\Psi(gg')$)
to implement (\ref{inducida}) and reduce in
this way the representation (\ref{regular}), which is highly reducible.
Left-invariant vector fields, once a set of local coordinates $\{g^i\}_{i=1}^{\rm dim G}$ is chosen, are easily computed
as
\[X^L_j(g)\equiv X^L_{g^j}(g)= \sum_{k=1}^{\rm dim G}(L_g^T(e))_j^k
\frac{\partial}{\partial g^k},\quad (L_g^T(e))_j^k=
\left.\frac{\partial (g g')^k}{\partial g'^j}\right|_{g'=e},\quad j,k=1,\dots,
{\rm dimG}\,.\]
Here $L_g^T(e)$ represents the tangent at the identity element of the left
translation on the group $L_g(g')=gg'$, and $X^L_j(g)$ is the left-invariant vector
field verifying $X^L_j(e)=\frac{\partial}{\partial g^j}$ (see, for instance,
\cite{Chevalley}).
These vector fields are left-invariant thanks to the chain rule, and they turn out to be the infinitesimal generators of the
right action of the group on ${\cal F}(G)$, $[U^R(g')\Psi](g)=\Psi(gg')$.
Where there is no confusion, we shall omit the dependence on the group element, and write $X^L_j$ for $X^L_j(g)$.

The set of left-invariant vector fields $\{X^L_j\}_{i=1}^{\rm dim G}$ is a basis of $T_g G$
closing a Lie algebra isomorphic to the Lie algebra ${\cal G}$ of $G$, and therefore they
constitute a realization of it acting on ${\cal F}(G)$.

Thus, if ${\cal P}$ is the Lie algebra of $P$, realized in terms of left-invariant vector fields, we impose a condition of the form
\be
X^L\Psi(g)=\alpha(X^L)\Psi(g)\,,\qquad \forall X^L\in {\cal P},
\label{polarizacion1} \ee
where now $\Psi\in C^1(G)$ and $\alpha$ constitutes a one-dimensional representation (character) of the
subalgebra ${\cal P}$, which is the infinitesimal character associated to $D_\alpha$.
Since the representation is
one-dimensional, the character $\alpha$ vanishes on the derived subalgebra of ${\cal P}$,
$\alpha([X^L,Y^L])=0\,,\forall X^L,Y^L\in {\cal P}$.  This means that the character $\alpha$ can be
non-trivial only on the quotient of ${\cal P}$ by its derived algebra,
${\cal A}\equiv {\cal P}/[{\cal P},{\cal P}]$ which is an Abelian algebra. In particular if ${\cal P}$ is
semisimple, the character $\alpha$ is trivial.

The subalgebra ${\cal P}$ will be called a polarization, more precisely, a first-order polarization\footnote{Originally, the definitions
of first-order and higher-order polarizations were associated with a central extension $\tilde{G}$ of $G$ (which defines a notion of
horizontality in the Lie algebra), since they were introduced in the framework of Geometric Quantization \cite{marmoIJMP,marmoBregenz,Posicion,marmo}.
Here we adapt the definition to an arbitrary group, and this results in a larger freedom in the choice of polarizations.}.

%The reduction of the representation is achieved by means of a polarization
%${\cal P}$. There are two kinds of polarizations: first-order or higher-order.

\begin{defn}
A {\bf first-order polarization}
is a proper subalgebra ${\cal P}$ of the Lie algebra ${\cal G}$ of $G$, realized
in terms of left-invariant vector fields.
%It must satisfy a maximality condition in order to define an irreducible representation.
\end{defn}

Usually we shall be interested in first-order polarizations  such that the representation
obtained after the polarization conditions are imposed, see eq. (\ref{repreX2}), is not one-dimensional.

The advantage of using polarization equations is two-fold. On the one hand
we have at our disposal the powerful machinery of partial differential
equations, Frobenius theorem, etc., and, on the other hand, we can
generalize them to account for higher-order differential operators. Thus,
we can define:

\begin{defn}
A {\bf higher-order polarization} is a proper subalgebra ${\cal P}^{HO}$ of the (left) universal enveloping
algebra ${\cal UG}$  of  ${\cal G}$.
%which also satisfies a maximality condition in order to define an irreducible representation.
\end{defn}

Higher-order polarization conditions are imposed in a similar way to (\ref{polarizacion1}):
\be
X^{HO}\Psi(g)=\alpha(X^{HO})\Psi(g)\,,\qquad \forall X^{HO}\in {\cal P}^{HO} \label{polarizacion2}
\ee
where $\Psi\in C^k(G)$, $k$ being the maximum degree of the differential
operators in ${\cal P}^{HO}$ (and can actually be infinity), and $\alpha$
is a 1-dimensional representation of ${\cal P}^{HO}$. Here
we also shall be interested in  higher-order polarization
leading to non one-dimensional representations.

%The notion of higher-order polarization
%can be generalized to be a (left) ideal of ${\cal UG}$ (that is, the
%commutator of two elements of ${\cal P}^{HO}$ does not need to be an
%element of
%${\cal P}^{HO}$, but rather an arbitrary element of
%${\cal UG}$ times an element of ${\cal P}^{HO}$), see \cite{marmo}, but here
%we shall restrict ourselves
%to the case where ${\cal P}^{HO}$ is a subalgebra.

An important fact about polarizations is that they are chosen in such a way that they are compatible with
the action of the group in (\ref{regular}). In fact, since left and right actions always commute in a
group, the conditions (\ref{polarizacion1}) (resp. (\ref{polarizacion2}))
imposed by the polarization equations (left-invariant vector
fields which generate the right finite group action) are preserved by the left action defined
by (\ref{regular}). This means that if ${\cal H}_\alpha\subset C^1(G)$ (resp. $C^k(G)$) is the space of solutions of
(\ref{polarizacion1}) (resp.(\ref{polarizacion2})), then
$U^L{\cal H}_\alpha\subset {\cal H}_\alpha$.
This is one of the main features of GAQ, in contrast to other approaches like Geometric
Quantization where the existence of polarizations compatible with a given group action
is not always guaranteed.

Conditions like (\ref{polarizacion2}) imposed by higher-order
polarizations have no counterpart at the level of finite group transformations. This is one of the main
advantages of using polarizations instead of equations like
(\ref{inducida}). Except for certain cases, denoted anomalous
\cite{auaral}, where specific representations require the use of
higher-order polarizations, almost all (unitary) irreducible
representations of $G$ can be obtained by means of first-order
polarizations. But even in these cases, higher-order polarizations are
useful for obtaining realizations in certain carrier spaces $X$ which are
not homogeneous spaces, that is, that are not of the form $X=G/H$ for any
$H$ subgroup of $G$.

\subsection{First-order polarizations}

Let us study in detail the case of first-order polarizations, in
particular the structure of the space ${\cal H}_\alpha$ of solutions of
(\ref{polarizacion1}) (or its finite group transformations counterpart (\ref{inducida})). This
is a system of linear homogeneous first-order partial differential
equations, and the form of its solutions can be described as follows.

\begin{prop}
The solutions of the equations (\ref{polarizacion1}) can be factorized as
$\Psi(g)=W_\alpha(g)\Psi_0(g)$, where $\Psi_0(g)$ is the general solution
of the $\alpha=0$ system (trivial representation of ${\cal P}$):
\be
X^L\Psi_0(g)=0\,,\qquad \forall X^L\in {\cal P}\, , \label{polarizacion-homogenea}
\ee
\end{prop}
\ni and $W_\alpha(g)$ is a particular solution of the $\alpha\not=0$ system.

\ni {\bf Proof:} Since vector fields are derivations, they satisfy
Leibnitz's rule, and therefore $W_\alpha(g)\Psi_0(g)$ is a solution of
(\ref{polarizacion1}) provided they satisfy the hypothesis of the proposition. To prove that any solution can be written in this
way, it suffices to show that an everywhere non vanishing solution
$W_\alpha(g)$ exist, and then $W^{-1}_\alpha(g)\Psi(g)$ satisfies the $\alpha=0$
system.

According to Levi-Malcev theorem (see \cite{Barut}), ${\cal P}$ can be decomposed as
the semidirect product of a solvable algebra ${\cal R}$ (the radical) by a semisimple algebra ${\cal S}$:\,
${\cal P}={\cal S} \circledS {\cal R}$. The 1-dim representation $\alpha$ is trivial on ${\cal S}$, and therefore
is determined by the radical ${\cal R}$. According to Lie's theorem (see \cite{Humphreys}), a set of operators closing a
solvable algebra can always be simultaneously diagonalized:
\be
X^L W_\alpha(g) = \alpha(X^L) W_\alpha(g)\,,\,\,\forall X^L\in {\cal R}\,.
\label{reducido}
\ee
where $\alpha$ is a  1-dim representation of ${\cal R}$. Since $\alpha$ is  trivial on the ideal  $[{\cal R},{\cal R}]$,
the solutions of (\ref{reducido}) are determined by the solutions for the Abelian algebra  ${\cal A}={\cal R}/[{\cal R},{\cal R}]$ which turns to equal
${\cal P}/[{\cal P},{\cal P}]$.
Let $A$ be the quotient group  $P/[P,P]$ where $[P,P]$ denotes the  commutator group of $P$ (note
that since we have supposed $P$ simply connected, $[P,P]$ is a closed subgroup,
see \cite{Chevalley}). Let
$W_\alpha^A(a)$ be a particular solution of the equations
\be
X^L W_\alpha^A = \bar{\alpha}(X^L) W_\alpha^A\,,\,\,\forall X^L\in {\cal A}\,,
\label{reducidoA}
\ee
which again exists by Lie's theorem, and where $\bar{\alpha}$ denotes the character of ${\cal A}$ which extends to the character $\alpha$ of ${\cal R}$. Using Frobenius theorem, we can choose local coordinates $\{a^i\}$ in $A$ in such a way that $X^L_i=\parcial{a^i}\,,\forall X^L_i\in {\cal A}$, and therefore
$W^A_\alpha(a)=e^{\sum_{i=1}^{\rm dim {\cal A}} \bar{\alpha}_i a^i}$, where we denote
$\bar{\alpha}_i=\bar{\alpha}(X^L_i)\,,X^L_i\in {\cal A}$. In particular $W^A_\alpha(a)$ can be chosen to be non-zero
everywhere. The solution $W^A_\alpha(a)$ on $A$ can be extended to a
particular solution $W_\alpha(g)$ of (\ref{reducido}), $W_\alpha(g)=e^{\sum_{i=1}^{\rm dim {\cal A}} \alpha_i a^i(g)}$, which satisfy the required properties.$\blacksquare$

%which we know it exists (and we know how to construct it) by Frobenius theorem
%(see, for instance \cite{Chevalley}) since ${\cal P}$ is a subalgebra, and $W_\alpha(g)$ is a
%particular solution of the $\alpha\not=0$ system.
%We are interested in constructing explicitly the function $W_\alpha(g)$, since it will used later.
%
%
%%
%
%
%we shall solve this equation on the quotient
%group  $A=P/[P,P]$, where $[P,P]$ denotes the derived or commutator group of $P$ (note
%that since we have supposed $P$ simply connected, $[P,P]$ is a closed subgroup,
%see \cite{Chevalley}).
%
%This system has always a solution, since a set of commuting operators can
%always be simultaneously diagonalized (this is true also for a set of operators closing a
%solvable algebra, see ) . In fact, using Frobenius theorem, we can choose local coordinates
%in $A$ in such a way that $X^L_i=\parcial{s^i}\,,\forall X^L_i\in {\cal A}$, we can write
%$W_\alpha(a)=e^{\sum_{i=1}^{\rm dim {\cal A}} \alpha_i s^i}$, where we denote
%$\alpha_i=\alpha(X^L_i)$. In particular $W_\alpha(a)$ can be chosen to be non-zero
%almost everywhere.

Let us study in detail the properties of the solutions $\Psi_0$ of
(\ref{polarizacion-homogenea}). The finite group transformations version of this equation is
written in terms of the right action $U^R$ of $G$ on ${\cal F}(G)$ as:
\be
[U^R(h)\Psi_0](g)\equiv\Psi_0(gh)=\Psi_0(g)\,,\forall h\in P
\label{pola-finita}, \ee
although in this case the differentiability requirement can be dropped, so
$\Psi_0\in {\cal F}(G)$. Let ${\cal H}_0\subset {\cal F}(G)$ be the space of solutions of the last
equation. If we denote by $X=G/P$, then we have the following
result:

\begin{prop}
There is an isomorphism between the space of solutions ${\cal H}_0$ of
(\ref{pola-finita}) and the space of complex (Borel)  functions ${\cal F}(X)$ on $X$.
\end{prop}

\ni {\bf Proof:} Let $\pi:\, G\rightarrow X=G/P$ be the canonical
projection. Then, for any $\phi\in {\cal F}(X)$ we can define a function
$\Psi_0\in {\cal F}(G)$ by $\Psi_0(g)=\phi(\pi(g))$. This function
verifies (\ref{pola-finita}) since
$\Psi_0(gh)=\phi(\pi(gh))=\phi(\pi(g))=\Psi_0(g)$.

On the other hand, if $\Psi_0\in {\cal H}_0$ is a solution of (\ref{pola-finita})
and $s:X\rightarrow G$ is a Borel section, we can define a function
$\phi\in {\cal F}(X)$ by $\phi(x)=\Psi_0(s(x))\,,\forall x\in X$. The
function $\phi$ is independent of the choice of Borel section, since if
$s'$ is another Borel section, then $s'(x)=s(x)h(x)$, with $h(x)\in
P\,,\forall x\in X$ (a gauge transformation). Then, using
(\ref{pola-finita}) we have that
$\phi'(x)=\Psi_0(s'(x))=\Psi_0(s(x)h(x))=\Psi_0(s(x))=\phi(x)$.
Therefore, $\phi$ is unique.\,$\blacksquare$

This result states that the space ${\cal H}_\alpha$ of solutions of
(\ref{polarizacion1}) (or rather the corresponding subspace of ${\cal
F}(G)$ of solutions of the finite group transformations counterpart (\ref{inducida})) is
isomorphic to ${\cal F}(X)$, since the function $W_\alpha(g)$ is fixed. We
can translate the action $[{\cal U}_\alpha(g)\Psi](g')$ of $g\in G$ on
$\Psi=W_\alpha \Psi_0\in {\cal F}(G)$  onto
$\Psi_0\in {\cal F}(X)$ in a suitable manner. First, let us define on ${\cal H}_0$ a
modified action:
\be
[{\cal U}_\alpha(g)\Psi_0](g')\equiv\ W_\alpha^{-1}(g') [U^L(g)(W_\alpha
\Psi_0)](g')= W_\alpha^{-1}(g')W_\alpha (g^{-1}g')\Psi_0(g^{-1}g')
\label{repreX1} \ee
where $\Psi_0\in {\cal H}_0$.
Now the function $\lambda_\alpha(g,g')\equiv W_\alpha^{-1}(g')W_\alpha (g^{-1}g')$ is really a function
on $G\times X$, since $\lambda_\alpha(g,g'h)= W_\alpha^{-1}(g'h)W_\alpha (g^{-1}g'h)=
D_{\alpha}(h)^{-1} W_\alpha^{-1}(g')D_\alpha(h)W_\alpha (g^{-1}g')=W_\alpha^{-1}(g')W_\alpha (g^{-1}g')
=\lambda_\alpha(g,g')$,
provided that $W_\alpha$ is a particular solution of (\ref{inducida}).

The function $\lambda_\alpha(g,x)$ is a multiplier and satisfies
 1-cocycle properties, see (\ref{1-cocycle}). This means that ${\cal
U}_\alpha(g)$ constitutes a representation of $G$ on ${\cal H}_0$.

The representation on ${\cal F}(X)$ is now defined, using a Borel section
$s:\,X\rightarrow G$, as:
\bea
{\cal U}_\alpha(g)\phi(x)&\equiv&\ W_\alpha^{-1}(s(x)) U^L(g)W_\alpha (s(x))\Psi_0(s(x)) \nn \\
&=& W_\alpha^{-1}(s(x))W_\alpha
(g^{-1}s(x))\Psi_0(g^{-1}s(x))=\lambda_\alpha(g,s(x))
\phi(\pi(g^{-1}s(x))) \label{repreX2} \\
&=& \lambda_\alpha(g,x) \phi(g^{-1} x) \nn
\eea
where $\Psi_0(g)\equiv \phi(\pi(g))$ and $gx$ is the natural action of $G$
on the coset space $X=G/P$. This representation does not depend on the
choice of ``representative" Borel section $s$, since $\lambda_\alpha$ is a function on $G\times X$.

The infinitesimal version of this representations can be easily computed.
Since right and left-invariant vector fields commute, the form  of the solutions will be
preserved by the action of right-invariant vector fields, i.e.
$X^R W_\alpha(g)\Psi_0(g)=W_\alpha(g)\Psi'_0(g)$. Thus, we can
define:
\be
{\cal X}_j \phi(x) \equiv  W^{-1}_\alpha(s(x))(iX^R_j) W_\alpha(s(x))\Psi_0(s(x))\,\qquad j=1,\ldots,
{\rm dim}(G)\,, \label{repreX3}
\ee
where the imaginary constant is introduced generally to obtain hermitian operators,
and where $\Psi_0(g)\equiv \phi(\pi(g))$.

To discuss the unitarity of the obtained representations, we need to
introduce a scalar product, that is, an invariant or quasi-invariant
measure on $X$. The existence of invariant measures on homogeneous spaces
is not always guarantied, but there always exist quasi-invariant measures
\cite{Mackey}. Given a quasi-invariant measure $d\mu(x)$ on $X$, we have
to introduce the corresponding Radon-Nikodym derivative
$\lambda(g,x)\equiv \frac{d\mu(gx)}{d\mu(x)}$.

In our approach we are able to construct quasi-invariant measures and the
corresponding Radon-Nikodym derivative in a easy way (see \cite{medida} for details).
The invariant or quasi-invariant measure is given by:
\be
d\mu = i_{X^L_p}i_{X^L_{p-1}} \cdots i_{X^L_1} d^Lg\,,\qquad X^L_1,X^L_2,\ldots,X^L_p \in {\cal P}
\label{medida}
\ee

\ni where $\{X^L_1,X^L_2,\ldots,X^L_p\}\,, p={\rm dimP}$, is a basis of ${\cal P}$ and
$i_{X}\omega$ denotes interior product (or contraction) of the vector field $X$ and the
differentiable form $\omega$.
As proved in \cite{medida}, $d\mu$ is an invariant measure on $X=G/P$ or can
be converted into a quasi-invariant one by multiplying it by an
appropriate function $\rho$ which satisfies an equation of the type
(\ref{inducida}). In fact, the function $\rho$ is of the form
$\rho(g)=|W_\alpha(g)|^2$ for a particular choice of the 1-dim
representation $D_\alpha$. The corresponding Radon-Nikodym derivative is
given by:
\be
\lambda(g,x)=\frac{\rho(gs(x))}{\rho(s(x))}
\ee
where $s:\,X\rightarrow G$ is any Borel section. The advantage of our
procedure is that the function $\rho$ is built in the solutions of the
equation (\ref{inducida}), in such a way that the quadratic form
$\bar{\Psi} \Psi'$ appearing in the scalar product leads to $|W_\alpha(g)|^2\bar{\Psi}_0(g)\Psi_0'(g)$ which
can be properly integrated on $X$ with respect to the measure $d\mu(x)$.

In this way, choosing appropriately the 1-dim representation $D_\alpha$, the representation
${\cal U}_\alpha$ can be made unitary acting on $L^2(X,|W_\alpha(g)|^2d\mu)$.

Note that if $X$ does not admit an invariant measure, then the
representation $D_\alpha$ must be non-unitary. Otherwise the function
$\rho(g)=|W_\alpha(g)|^2$ belongs to ${\cal H}_0$ and the Radon-Nikodym
derivative is trivial or it is a 1-coboundary (see below).

Choosing different particular solutions $W_\alpha$ and $W_\alpha'$ of
(\ref{reducido}) (or (\ref{inducida})) leads to (unitarily) equivalent representations.
We only need to take into account that $W_\alpha'(g)=
W_\alpha(g)\rho(g)$, with $\rho\in {\cal H}_0$. Then, if $s:\,X\rightarrow G$ is a Borel section:
\be
\lambda_\alpha'(g,s(x))=W_\alpha'{}^{-1}(s(x))W_\alpha'(g^{-1}s(x))=
W_\alpha^{-1}(s(x))W_\alpha(g^{-1}s(x))\frac{\rho(g^{-1}s(x))}{\rho(s(x))}
\ee

Now the function $\lambda(g,x)=\frac{\rho(g^{-1}s(x))}{\rho(s(x))}$ is a
trivial 1-coycle, i.e., a 1-coboundary generated by the 0-cocycle
$\rho(s(x))$. This means that the multipliers $\lambda_\alpha$ and
$\lambda_\alpha'$ are equivalent (see \cite{Bargmann}) and therefore the
defined representations ${\cal U}_\alpha$ and ${\cal U}_\alpha\,'$ are
(unitarily) equivalent.

Let us discuss on the irreducibility of the representations obtained by
first order polarizations.

\begin{defn}
A unitary representation of $G$ is said to be {\bf quasi-irreducible} if
the set of operators commuting with the representation, up to multiplicative constants, is discrete.
\end{defn}

A unitary quasi-irreducible representation decomposes into a discrete sum of irreducible representations.

\begin{defn}
A first-order polarization ${\cal P}$ is said to be {\bf full} (or {\bf regular}) if the representation obtained after the polarization conditions are imposed, eq. (\ref{repreX2}),
is irreducible or quasi-irreducible.
\end{defn}

In the case of a full polarization leading to a quasi-irreducible representation, we shall need to resort to a non-connected
subgroup $P\subset G$, or even to discrete operators external to the group
to fully reduce the representation (see for instance Sec. \ref{afinR}, after eq. (\ref{hijaAfin}), and Sec. \ref{contser}, after eq. (\ref{lambda})).

A non-full first-order polarization ${\cal P}$  will lead to a reducible
representation. The (quasi-) irreducibility will be accomplished by enlarging ${\cal P}$ with elements either of the Lie Algebra, thus leading
to a larger full first-order polarization, or of the enveloping algebra ${\cal U}({\cal G})$, thus leading to a higher-order polarization.

Another problem is that the polarization subgroup $P$ can be multiply
connected, even though we have supposed that $G$ is simply connected (think, for instance,
in $U(1)\subset SU(2)$). In this case, not all infinitesimal characters
$\alpha$ of ${\cal P}$ are exponentiable to a character $D_\alpha$ of $P$,
and some restrictions appear (like half-integer values of spin for
$SU(2)$).

In certain circumstances, it is convenient to resort to complex
polarizations ${\cal P}\subset {\cal G}^c$ in order to obtain
representations on carrier spaces with a K\"ahler structure (as, for
instance, for the discrete series of representations for semisimple Lie
groups). In this case, the use of polarization subalgebras avoids the need
of complexifying the group, only the polarization is complex and the
carrier space of the representation is $X=G/H$ where $H$ is a subgroup
whose Lie algebra is a real subalgebra of ${\cal P}$. The space $X$ admits
a K\"ahler structure, and in this case the function
$W_\alpha(g)$ is related to the K\"ahler potential.

Finally, let us comment that we have chosen to impose polarizations with
finite right group actions $U^R$ (i.e. with left-invariant vector fields $X^L$)
and group actions with finite left group actions $U^L$. Nevertheless, left and
right can be interchanged at computational convenience, and the
representations obtained are (unitarily) equivalent. Thus, we shall choose
the more convenient option along the examples of this paper.

\subsection{Higher-order polarizations}

Although higher-order polarizations have been widely used in GAQ, there
are not many conclusive results describing the general properties of the
solution spaces and the unitarity of the representations (as the existence
of invariant measures, multipliers, etc.). However we shall give an operative
description of how they work (see for instance,
\cite{marmoIJMP,marmoBregenz,Posicion,marmo}).

The first difference with first-order polarizations lies in the fact that
Leibnitz's rule does not apply for higher-order differential operators,
therefore factorizations like the one used in the previous subsection are
not valid. That is, the general solution of (\ref{polarizacion2}) cannot
be written as $\Psi(g)\neq W_\alpha(g) \Phi(g)$. This implies, in
particular, that the space of solutions ${\cal H}_\alpha \subset C^k(G)$
is not isomorphic to $C^k(X)$, with $X=G/H$, for any subgroup $H$ of $G$.
That is, the carrier space of the representation will not be an
homogeneous space.

In many examples studied,  the general solution of
(\ref{polarizacion2}) turns out to have  the form:
\be
\Psi(g) = \hat{W}_\alpha(g) \Phi(x) \label{solucionHO}
\ee

\ni where $\hat{W}_\alpha(g)$ is an infinite-order (pseudo-)differential operator
acting on $\Phi\in C^\infty(X)$ and $X$ is a non-homogeneous submanifold of
$G$. With the appropriate coordinates, $\hat{W}_\alpha(g)$ can be written
as:
\be
\hat{W}_\alpha(g) = e^{\sum_i s^i(g)\hat{O}_i(x)},
\ee
where $i$ runs from $i=1,\ldots,dim({\cal P}^{HO})$ and $\hat{O}_i(x)$ denote a set of
higher-order differential operators acting on $C^\infty(X)$ related to the higher-order
polarization. An example of this will be given in Sec. \ref{sl2rRmas}, eqs.
(\ref{solHOP1}) and (\ref{solHOP2}).

Once the solutions of the higher-order polarization have
been found, the representation (\ref{regular}) can be reduced to the space
$X$. In this case, the finite group action cannot be computed, since it is not
transitive, but we can compute the infinitesimal action by means of
right-invariant vector fields. Since right and left-invariant vector
fields commute, the form (\ref{solucionHO}) of the solutions will be
preserved by the action of right-invariant vector fields, i.e.
$X^R \hat{W}_\alpha(g)\Phi(x)=\hat{W}_\alpha(g)\Phi'(x)$.
Thus, $\hat{W}^{-1}_\alpha(g)X^R \hat{W}_\alpha(g)\Phi(x)$ is a well-defined operator
on $C^\infty(X)$, and we can define:
\be
{\cal X}_j \Phi(x) \equiv  \hat{W}^{-1}_\alpha(g)(iX^R_j)\hat{W}_\alpha(g)\Phi(x)\,\qquad j=1,\ldots,
{\rm dim}(G)\,.
\ee

This defines a representation of $G$ on $C^\infty(X)$ by means of
pseudo-differential operators (see eq. (\ref{solHOP3}) for the above-mentioned
example). This representation can be converted to a
unitary representation on $L^2(X,d\mu)$ if an appropriate measure $d\mu$
is chosen (see Sec. \ref{sl2rRmas}, after eq. (\ref{op1})). However, there are no conclusive results concerning
the existence of invariant or quasi-invariants measures for higher-order
polarizations (even these notions should be properly defined for
non-homogeneous spaces).

In some cases, the operators ${\cal X}_i$ should be modified in order to
obtain a unitary representation, in a process that can be seen as the
generalization for higher-order polarizations of the Radon-Nikodym
derivative (see \cite{oscipert}).

In most cases, a higher-order polarization is obtained by adding
higher-order operators to a first order, non-full, polarization.
In particular, Casimir operators are very useful for this purpose, since they are
central in the enveloping algebra and therefore can be added to any
polarization (first or higher-order).

As far as irreducibility is concerned, the same considerations as in
the case of first-order polarizations apply.

In certain cases, (real) higher-order polarizations can be used
instead of complex polarizations. In this case, we either obtain the
realization in the same K\"ahler manifold
 (see for instance Sec. \ref{sl2rCmas}), or a representation on a
 non-homogeneous manifold is obtained, like in the case of the harmonic
 oscillator in configuration space, either non-relativistic or
 relativistic (see \cite{oscilata}).

 \subsubsection{Second-order polarizations}

 The case of second-order polarizations merits special attention, since some results from the theory
 of partial differential equations can be applied to this case. Also, all higher-order polarizations
 appearing in this work are of second order type.

 Second-order partial differential equations can be classified into three
 types, according to the eigenvalues of the coefficient matrix of the second order terms. They can be
 elliptic,  if all eigenvalues are positive (or all negative), hyperbolic, if one of the eigenvalues is
 negative an the rest are positive, and parabolic if some eigenvalues are zero.

 For the case of $SL(2,\mathbb R)$, the Casimir operator, see eq. (\ref{casimirsl2r}), is a hyperbolic
 second-order differential operator, as can be checked by direct computation, or using the fact that the invariant Killing metric
 has signature $(+,+,-)$. Here, a second-order polarization consists of the
 Casimir and a left (or right) invariant vector field. We shall first solve the (first-order)
 equation associated with the vector field, and substitute the solution into
 the Casimir equation. This produces a second-order partial differential
 equation (with two independent variables) that can be hyperbolic,
 elliptic or parabolic.

 The elliptic equation in this case is solved by factorization, using
 complex variables. The factorization of the second-order differential operator
 implies that the polarization equation is reduced to a first-order equation, and therefore
 the results of first-order polarization apply. An example of this can be found in Section \ref{sl2rCmas}.
 Also, in this section other possible solutions of the elliptic equation are considered.

 The parabolic equation appears in Section \ref{sl2rRmas}, and in this case it has the form of a
 Schr\"odinger-like equation (Parabolic equations generally lead  to
 Heat-type equations, but since we are considering unitary
 representations, the equations must imply a unitary evolution). This
  Schr\"odinger-like equation is solved by the method explained previously, leading to a representation
  that is genuinely of higher-order type.

  The hyperbolic equation, which appears in Sec. \ref{polahyperbolic}, can be solved using a factorization
method. In this case a representation with support in $\mathbb S^1$ is obtained, leading to the
continuous series of representations, obtaining the same results as in Sec. \ref{contser} with a
first-order polarization.

\subsection{Polarization-changing operators}
\label{PCO}

Not all polarizations lead to different, nonequivalent representations. In
many cases, two different polarizations (even of different type, one
first-order and the other higher-order) lead to equivalent representations
which are however realized in different carrier spaces. In this case we
would like to compute the unitary operator which relates both
representations. In this context, we shall call them polarization-changing
operators, and they can be constructed as follows.

Let us denote by ${\cal P}_X$ the polarization (first or higher-order) leading to the
representation ${\cal U}_X$ (or series of representations) on the Hilbert space $L^2(X,dx)$,
where $dx$ is an appropriate measure on $X$ that makes the representation unitary.
Then, if the polarizations ${\cal P}_X$ and ${\cal P}_Y$ lead to equivalent representations ${\cal U}_X$
and ${\cal U}_Y$, there exist a unitary isomorphism $V:\,L^2(X,dx)\rightarrow L^2(Y,dy)$ such that
${\cal U}_Y V = V {\cal U}_X$. The unitary operator $V$ will be in general an integral operator
characterized by a kernel
\be
\tilde{\phi}(y) \equiv V \phi(x) = \int_X K(x,y) \phi(x) dx
\ee

For convenience, we shall use Dirac's bra-ket notation, and write
$\phi(x)\equiv \langle x|\phi\rangle\in L^2(X,dx)$ and $\tilde{\phi}(y)\equiv \langle y|\phi\rangle\in L^2(Y,dy)$.\footnote{In order to avoid
confusion, we shall denote with $\phi(x)$ and $\tilde{\phi}(y)$ the realization of the same ``state" $|\phi\rangle$ in different Hilbert spaces.}
In Dirac's notation bras and kets can be separated and have meaning by themselves:
kets are vectors in an abstract Hilbert space and bras belong to its dual.

Completeness relations (resolutions of unity) and reproducing kernels
\cite{Gazeau} adopt a very intuitive form in Dirac's notation. For
instance, we can formally write:
\begin{equation}
\tilde{\phi}(y)=\langle y|\phi\rangle = \int_X dx\langle y|x\rangle \langle x|\phi\rangle
=\int_X dx\langle y|x\rangle \phi(x)
\label{PCO1}
\end{equation}
where the completeness relation $I=\int_X dx|x\rangle \langle x|$
has been introduced. The factor $\langle y|x\rangle$ in the
integral is the kernel for the integral transform (\ref{PCO1})
which relates the representation on $L^2(X,dx)$ to the one in
$L^2(Y,dy)$. To compute this integral kernel we can proceed as
follows. Search for a basis of states $\{|n\rangle\}$ of the
Hilbert space, where the index $n$ is either discrete, continuous
or of both types. This basis should be easily computed in both
Hilbert spaces $L^2(X,dx)$ and $L^2(Y,dy)$. Then introduce the
completeness relation $I=\int |n\rangle\langle n|d\mu(n)$:
\be \langle y|x\rangle = \int \langle y|n\rangle \langle
n|x\rangle d\mu(n)\ee
where $d\mu$ accounts for a discrete and/or continuous measure.
For the basis $\{|n\rangle\}$ we generally use the eigenfunctions
of some of the generators of the infinitesimal action of the group
$G$ on $X$, which are self-adjoint and therefore their spectrum
are complete in the Hilbert space.

Examples of polarization-changing operators are given in
\cite{RBT,oscilata}, where a relativistic counterpart of the Bargmann
transform relating configuration space and Bargmann-Fock space for the
harmonic oscillator is given, and in Sec. \ref{afinRmas} and
\ref{sl2rCmas} where the Fourier and Laplace transform are derived. See
also \cite{RadiacionVacio} for more examples.

\section{Wavelets for the affine group}
\label{afin}

%We shall illustrate the procedure developed previously with the case of
Let us consider the affine group in one dimension, $SIM(1)$, which will serve us as an introduction for
studying the most
interesting case of $SL(2,\mathbb{R})$ which will be considered in the next section.

The affine group $G=SIM(1)=\mathbb R\rtimes \mathbb R^+=\{ g=(b,a)/\,b\in\mathbb{R}, a\in \mathbb R^+\}$ of translations
and dilations in one dimension is given by the following group law
($g''=g'g$):
\begin{equation}
\begin{array}{rcl}
a'' &=& a'a \\
b'' &=& b + ab'
\end{array}
\end{equation}

This group law can be obtained from the upper triangular matrices with determinant
one and positive diagonal, which admit the following Iwasawa $KAN$ decomposition
(with K trivial in this case):
\begin{equation}
g=\left(
\begin{array}{cc}
 \frac{1}{\sqrt{a}} &  0\\
0 &\sqrt{a}
\end{array}\right)
\left(
\begin{array}{cc}
 1 &  b\\
0 &1
\end{array}
\right)
=\left(
\begin{array}{cc}
 \frac{1}{\sqrt{a}} & \frac{b}{\sqrt{a}} \\
0 &\sqrt{a}
\end{array}\right)
\end{equation}

Left and right-invariant vector fields are immediately computed from the group law:
\begin{equation}
\begin{array}{rcl}
X^L_a &=& a\parcial{a} + b\parcial{b}\\
X^L_b &=& \parcial{b}
\end{array} \qquad
\begin{array}{rcl}
X^R_a &=& a\parcial{a}\\
X^R_b &=& a\parcial{b}
\end{array}
\ee

Left and right-invariant Haar measures are also easily obtained:
\begin{eqnarray}
d^Lg=\frac{1}{a}da\wedge db \\
d^Rg=\frac{1}{a^2}da\wedge db
\end{eqnarray}
Since they are different, this group is not unimodular. Therefore we should take
care of choosing the proper measure in each case.

Selecting the appropriate polarization, different (equivalent in this case)
representations can be obtained for the affine group.

\subsection{Representation on $L^2(\mathbb{R})$}
\label{afinR}

Let us consider the polarization subalgebra ${\cal P}_{\mathbb R}=<X^R_a>$
(in this case the right polarization is the most convenient). Therefore,
the representation space is the subspace of ${\cal F}(G)$ of functions
$\Psi(g)$ satisfying
$X^R_a \Psi = \alpha \Psi\,,\alpha\in\mathbb C$. The solutions have the form:
\begin{equation}
\Psi(a,b)=W_\alpha(a,b) \phi(b)=a^\alpha \phi(b)   \,\,,\qquad \phi\in L^2(\mathbb{R},d\mu)
\end{equation}

Later we shall determine the values of $\alpha$ and the measure $d\mu$ in order
to have a unitary representation.

The representation is realized by the operators $X^L_a$ and $X^L_b$ acting
on this space of functions. Since the factor $a^\alpha$ is common to all
functions, it is convenient to subtract it and consider the realization in
$L^2(\mathbb{R},d\mu)$. Therefore, as in eq. (\ref{repreX3}), we define the operators:
\be
\ba{rcl}
{\cal X}_a&\equiv & a^{-\alpha}(iX^L_a)a^{\alpha} = ib\parcial{b}+i\alpha \\
{\cal X}_b&\equiv & a^{-\alpha}(iX^L_b)a^{\alpha} = i\parcial{b}
\ea
\ee

Following the general construction, the measure $d\mu$ can be computed as
$d\mu=i_{X^R_a}d^Rg=\frac{1}{a}db$. This is not a well-defined measure on
$\mathbb{R}$, and this reflects the fact that the quotient space
$G/A\approx \mathbb{R}$ does not admit an invariant measure. Therefore we
should look for quasi-invariant measures and introduce the appropriate
Radon-Nikodym derivative. In this simple case, this amounts to
restrict the allowed values of $\alpha$ to
$\alpha=\frac{1}{2}+is,\,,\,s\in\mathbb{R}$. In this way, if
$\Psi=a^{\frac{1}{2}+is}\phi(b)$ and $\Psi'=a^{\frac{1}{2}+is}\phi'(b)$ are two
functions, the scalar product is:
\begin{equation}
\langle \Psi,\Psi'\rangle = \int_{-\infty}^\infty \frac{db}{a} a^{\frac{1}{2}-is}
\bar{\phi}(b)a^{\frac{1}{2}+is}\phi'(b)= \int_{-\infty}^\infty db \bar{\phi}(b)\phi'(b)
\end{equation}
from which we deduce that the representation is unitary with the measure
$d\mu=db$. The Radon-Nikodym derivative appears automatically as the
factor $a^{ \frac{1}{2}}$ in the functions $\Psi$. We shall leave for the
next subsection the discussion of the irreducibility.

The finite group action for this representation is given by the right action. When restricted
to $L^2(\mathbb{R})$ is:
\be
{\cal U}_\alpha(g)\equiv a^{-\alpha} {\cal U}^R(g^{-1}) a^{\alpha}
\ee
in such a way that
\be
{\cal U}_\alpha(a',b')\phi(b)=(a')^{-\alpha}\phi(\frac{b-b'}{a'})
\ee

The representations ${\cal U}_\alpha$ with $\alpha=\frac{1}{2}+is\,,s\neq
0$ are unitarily equivalent to the representation ${\cal U}_\frac{1}{2}$,
with $s=0$, through the intertwining operator $T_s=a^{is}$, ${\cal
U}_{\frac{1}{2}+is}T_s = T_s {\cal U}_{\frac{1}{2}}$. Therefore, we shall
always work with $\alpha=\frac{1}{2}$, and denote
${\cal U}\equiv{\cal U}_{\frac{1}{2}}$.

The machinery of wavelets analysis proceeds now defining, for an admissible function
$\gamma\in L^2(\mathbb{R})$:
\be
\gamma_{b',a'}(b)={\cal
U}(a',b')\gamma(b)=(a')^{-\frac{1}{2}}\gamma(\frac{b-b'}{a'})
\label{hijareal}
\ee

\subsection{Representation on $L^2(\mathbb{R}^+,\frac{da}{a})$}
\label{afinRmas}

Let us consider now the polarization ${\cal P}_{\mathbb R^+}=<X^L_b>$ (the
left polarization is the best choice now). Therefore, the representation
space is the subspace of ${\cal F}(G)$ of functions
$\Psi(g)$ satisfying $X^L_b \Psi = \beta \Psi\,,\beta\in\mathbb C$. The solutions have the
form:
\begin{equation}
\Psi(a,b)=W_\beta(a,b)\phi(a)=e^{\beta b} \phi(a)   \,\,,\qquad \phi\in L^2(\mathbb{R}^+,d\mu)
\end{equation}

As before, we shall determine later the values of the complex parameter
$\beta$ and the measure
$d\mu$ in order to have a unitary representation.

The representation is realized, in this case, by the operators $X^R_a$ and
$X^R_b$ acting on this space of functions. Since the factor $e^{\beta b}$
is common to all functions, it is convenient to subtract it and consider
the realization in $L^2(\mathbb{R}^+,d\mu)$. Therefore we define the
operators:
\be
\ba{rcl}
{\cal X}_a&\equiv & e^{-\beta b}(iX^R_a)e^{\beta b} = ia\parcial{a} \\
{\cal X}_b&\equiv & e^{-\beta b}(iX^R_b)e^{\beta b}= i\beta a
\ea
\ee

Following the general construction, the measure $d\mu$ can be computed as
$d\mu=i_{X^L_b}d^Lg=\frac{da}{a}$. In this case, we obtain a well defined measure
on $\mathbb{R}^+$ and no Radon-Nikodym derivative is needed. The
representation is unitary if $\rho\equiv i\beta \in \mathbb{R}$. If
$\beta\neq 0$, we can define
$r\equiv \rho a$ in such a way that
\be
\ba{rcl} {\cal X}_a&= &i r\parcial{r} \\ {\cal X}_b&=& r\ea \ee
and $d\mu=\frac{dr}{|r|}$. This representation is unitary and irreducible
on
$L^2(\mathbb{R}^+,\frac{dr}{r})$ if $\rho>0$ and on $L^2(\mathbb{R}^-,-\frac{dr}{r})$
if $\rho<0$.

If $\beta=0$, it turns out that ${\cal X}_b=0$, and the representation is no longer
irreducible. It rather decomposes in the direct integral of the eigenspaces of the
operator ${\cal X}_a$.

The case $\beta=0$ will be discarded, and the nonequivalent cases are
given by $\rho=\pm 1$. We shall focus on the case $\rho=1$, the other can
be dealt in an analogous way. The finite group action is given by the left
action, which, when restricted to $L^2(\mathbb{R}^+,\frac{da}{a})$, is
given by:
\be
{\cal U}(g)\equiv e^{ib} {\cal U}^L(g) e^{-ib}
\ee
in such a way that
\be
{\cal U}(a',b')\phi(a)=e^{-iab'}\phi(a'a) \label{hijaAfin}
\ee

We wonder now what is the relation between the two unitary irreducible representations here
obtained on $\mathbb{R}^+$ and $\mathbb{R}^-$, and the unitary representation on $\mathbb{R}$
obtained in the previous subsection. The answer is provided by the Fourier transform, which
tells us that $L^2(\mathbb{R},db)$ is reducible under the action of the affine group and decomposes
as $L^2(\mathbb{R})=H^2_+(\mathbb{R})\oplus H^2_-(\mathbb{R})$, where $H^2_+(\mathbb{R})$ and
$H^2_-(\mathbb{R})$ are the subspaces of progressive and
regressive wavelets, respectively \cite{Holschneider}. Therefore, the
Fourier transform is a unitary isomorphism between $H^2_+(\mathbb{R})$ and
$L^2(\mathbb{R}^+,\frac{da}{a})$ and between $H^2_-(\mathbb{R})$ and
$L^2(\mathbb{R}^-,-\frac{da}{a})$.

Let us construct, in an explicit way, this unitary isomorphism. This can
be achieved by means of the ``polarization-changing operators" introduced
in Sec. \ref{PCO}. Using Dirac's notation, we can write formally:
\begin{equation}
\tilde{\phi}(b)=\langle b|\phi\rangle = \int_0^\infty \frac{da}{a}\langle b|a\rangle \langle a|\phi\rangle
=\int_0^\infty \frac{da}{a}\langle b|a\rangle \phi(a)
\label{Fourier1}
\end{equation}
where the completeness relation $I=\int_0^\infty \frac{da}{a}|a\rangle
\langle a|$ has been introduced. The factor $\langle b|a\rangle$ in the
integral is the kernel for the integral transform (\ref{Fourier1}) which
relates the representation on $L^2(\mathbb{R})$ with the one on
$L^2(\mathbb{R}^+,\frac{da}{a})$. To compute this integral kernel we can
proceed as explained in Sec. \ref{PCO}. For the basis $\{|n\rangle\}$ we
can use the eigenfunctions of the ``translation'' operator ${\cal X}_b$,
$\{|\beta\rangle\}\,,\,\beta\in  \mathbb{R}$,  which in $L^2(\mathbb{R})$ are
$\langle b|\beta\rangle=e^{i\beta b}$, whereas in $L^2(\mathbb{R}^+,\frac{da}{a})$ write
$\langle a|\beta\rangle=\delta(a-\beta)$ (in this case we must consider more general Hilbert spaces
to account for distributions). Note that $\langle a|\beta\rangle$ is identically zero for $\beta< 0$
(something similar happens in $L^2(\mathbb{R}^-,-\frac{da}{a})$ for $\beta>0$), and this is the
clue for the reducibility of the representation of the affine group on  $L^2(\mathbb{R})$. Therefore, we can write:
\be
\langle b|a\rangle = \int_{-\infty}^\infty d\beta \langle b|\beta\rangle \langle \beta|a\rangle
= \int_0^\infty d\beta e^{i \beta b}\delta(a-\beta)=e^{i ab}
\ee

Thus, equation (\ref{Fourier1}) can be written as:
\begin{equation}
\tilde{\phi}(b)=\int_0^\infty \frac{da}{a}e^{iab} \phi(a)
\label{Fourier2}
\end{equation}

Equation (\ref{Fourier2}) would be the analog of the inverse Fourier transform between
$L^2(\mathbb{R}^+,\frac{da}{a})$ and $L^2(\mathbb{R})$. More precisely, this is the composition of the
ordinary inverse Fourier transform between $L^2(\mathbb{R}^+)$ and $L^2(\mathbb{R})$
(in fact $H^2_+(\mathbb{R})$) and the unitary isomorphism between $L^2(\mathbb{R}^+)$ and
$L^2(\mathbb{R}^+,\frac{da}{a})$.

The direct Fourier transform is given by the adjoint operator, which has
as integral kernel the conjugate $\overline{\langle b|a\rangle}=\langle
a|b\rangle=e^{-iab}$.

In an analogous manner, for $a<0$ we have the corresponding integral transform between $L^2(\mathbb{R})$
(or rather $H^2_-(\mathbb{R})$) and $L^2(\mathbb{R}^-,-\frac{da}{a})$.

\subsection{Representation on the right half plane}

We can still obtain another representation of the affine group, this time on the right half complex plane
$\mathbb C^+\equiv \{w\in\mathbb{C}/ \Re(w)>0\}$. For this purpose the complex polarization
${\cal P}_{\mathbb C^+}=<X^R_a+iX^R_b>$ can be used.

To obtain the solutions of this polarization it is better to exploit the complex structure of the manifold
of the affine group, and to perform the change of variable:
\be
\ba{rcl} w&=&a+ib\\ \bar{w}&=&a-ib \ea \ee

In the new variables, the solutions to the polarization $(X^R_a+iX^R_b)\Psi=\nu\Psi\,,\nu\in\mathbb C$ are of the form:
\be
\Psi(w,\bar{w})=W_\nu(w,\bar{w})\phi(w)= (\Re(w))^{\frac{\nu}{2}}\phi(w) \ee

That is, the carrier space of the representation is, apart from a common
factor $(\Re(w))^{\frac{\nu}{2}}$, the space of holomorphic functions over
$\mathbb C^+$.

The operators realizing the representation are left invariant vector
fields acting on this space of functions. If we subtract the common
dependence in $(\Re(w))^{\frac{\nu}{2}}$, the resulting operators are:
\be
\ba{rcl}
{\cal X}_a&\equiv& (\Re(w))^{-\frac{\nu}{2}}(iX^L_a)(\Re(w))^{\frac{\nu}{2}}=iw\parcial{w}+i\frac{\nu}{2}\\
{\cal X}_b&\equiv& (\Re(w))^{-\frac{\nu}{2}}(iX^L_b)(\Re(w))^{\frac{\nu}{2}}=-\parcial{w}
\ea
\ee

The measure in this case is the very right-invariant Haar measure, written
in these coordinates, $d\mu=(\Re(w))^{-2}dw\wedge d\bar{w}$. The
representation obtained is irreducible and unitary for $\nu\in
\mathbb{R}$.

The finite group action, restricted to the holomorphic part of the functions, is
given by the right action:
\be
{\cal U}_\nu\equiv (\Re(w))^{-\frac{\nu}{2}}{\cal
U}^R(\Re(w))^{\frac{\nu}{2}} \ee
having the form:
\be
{\cal U}_\nu(a',b')\phi(w)=(a')^{-\frac{\nu}{2}}\phi(\frac{w-ib'}{a'}) \ee

All these representations for $\nu\in \mathbb{R}$ are equivalent to the
case $\nu=0$.

The polarization-changing operators that connect this representation with
others previously obtained could also be computed. It is particularly
interesting the one relating this representation to the one on
$L^2(\mathbb{R}^+,\frac{da}{a})$, which is the Laplace transform. However,
we shall postpone its explicit computation until we study the
representations of the $SL(2,\mathbb{R})$ group, since there, with the
help of the generating function for the generalized Laguerre polynomial,
the computation will be straightforward (see Sec. \ref{sl2rRmas}).

\section{CWT on $\mathbb{R}$ and $\mathbb S^1$ from $G=SL(2,\mathbb{R})$\label{sl2r}}

In this Section we shall identify the group of affine transformations of
$\mathbb S^1$ and
$\mathbb{R}$, namely $SL(2,\mathbb{R})$, and we shall derive the CWT on both spaces
in a unified manner, based on the construction of general CS associated to
square integrable representations (modulo a subgroup) of
$SL(2,\mathbb{R})$ obtained from different (first- or higher-order) polarization subalgebras.
 We shall see that both transforms coincide in the
Euclidean limit.

Our main ingredient will be the composition group law $g''=g'g$ for
$SL(2,\mathbb R)$. In order to compute it, let us use the Iwasawa
decomposition (see, for instance, \cite{Barut}) to parameterize the
$SL(2,\mathbb{R})$ group:
\begin{equation}
g=\left(\begin{array}{cc} \cos\theta &-\sin\theta \\ \sin\theta&\cos\theta \end{array}\right)
 \left(\begin{array}{cc} \frac{1}{\sqrt{a}}    &0 \\0& \sqrt{a} \end{array}\right)
\left(\begin{array}{cc} 1 &b \\0&1 \end{array}\right)
   =
   \left(\begin{array}{cc}
 \frac{\cos \theta }
   {{\sqrt{a}}} & \frac{b\,\cos\theta}
    {{\sqrt{a}}} - {\sqrt{a}}\,\sin \theta  \cr
    \frac{\sin \theta }{{\sqrt{a}}} &
   {\sqrt{a}}\,\cos \theta  +
   \frac{b\,\sin \theta }{{\sqrt{a}}}
   \end{array}\right)
 \end{equation}
where $a\in\mathbb R^+,\,b\in\mathbb R$ and $\theta\in(-\pi,\pi]$.

Using matrix multiplication we can derive the composition law for the
parameters $\{a,b,\theta\}$, in the form $g''=g'g$, with:
\begin{eqnarray}
a'' &=& \frac{a\,a'}
  {{\cos^2 \theta}+
    \left( {a'}^2 + {b'}^2
       \right) \,{\sin^2 \theta } +
    b'\,\sin 2\,\theta} \nonumber \\
b'' &=&  \frac{\left( b + a\,b' \right) \,
     {\cos^2 \theta } +
    \left( 2\,b\,b' +
       a\,\left( -1 + {a'}^2 +
          {b'}^2 \right)  \right) \,
     \cos \theta\,\sin \theta  +
    \left( {a'}^2\,b +
       b'\,
        \left( -a + b\,b' \right)  \right) \,
      {\sin^2 \theta}}{{\cos^2 \theta } +
    \left( {a'}^2 + {b'}^2
       \right) \,{\sin^2 \theta} +
    b'\,\sin 2\,\theta }
 \nonumber\\
\tan\theta'' &=& \frac{a'\,
      \cos \theta'\,\sin \theta  +
     \left( \cos \theta  +
        b'\,\sin \theta  \right) \,
      \sin \theta'}{\cos \theta \,
      \cos \theta' +
     \sin \theta \,\left( b'\,
         \cos \theta' -
        a'\,\sin \theta'
        \right) }\label{grouplaw}
\end{eqnarray}

Left and right-invariant vector fields can be easily derived from the group law:
\begin{eqnarray}
X^{L}_a &=& a\,\parcial{a} + b\,\parcial{b}\nonumber \\
 X^{L}_b &=& \parcial{b} \\
X^{L}_\theta &=& - 2\,a\,b\,\parcial{a}+ \left( a^2 - b^2 -1\right) \,\parcial{b}+
a\,\parcial{\theta}   \nonumber
\end{eqnarray}
\begin{eqnarray}
X^{R}_a &=&  a\,\cos 2\,\theta \,\parcial{a} + a\,\sin 2\,\theta \,\parcial{b} +
  \cos \theta \,\sin \theta \,\parcial{\theta} \nonumber \\
X^{R}_b &=& - a\,\sin 2\,\theta \,\parcial{a} +
  a\,\cos 2\,\theta \,\parcial{b} -\sin^2 \theta \,\parcial{\theta}\\
X^{R}_\theta &=& \parcial{\theta}   \nonumber
\end{eqnarray}

The commutation relations for left-invariant vector fields (for right-invariant ones
they are the same except for the sign) is:
\begin{eqnarray}
{}[X^{L}_a,X^{L}_b]&=&-X^{L}_b\nonumber \\ {}[X^{L}_a,X^{L}_\theta]&=&2
X^{L}_b + X^{L}_\theta \label{sl2rcom}\\ {}[X^{L}_b,X^{L}_\theta]&=&
-2X^{L}_a \nonumber
\end{eqnarray}

The group $SL(2,\mathbb{R})$, being a simple group of rank one, has only one quadratic Casimir \cite{Barut}. It can be checked by direct computation
that the  following quadratic operator commutes with all left and right invariant vector fields, and therefore is the Casimir operator:
\begin{eqnarray}
\hat{C} &=& (X^{L}_a)^{2} + (X^{L}_b)^{2} +\frac{1}{2}\left(X^{L}_b
X^{L}_\theta +X^{L}_\theta X^{L}_b\right) \nn \\
&=&   a^2 \left( \frac{\partial^2}{\partial a^2} + \frac{\partial^2}{\partial
       b^2}\right) +  a \frac{\partial^2}{\partial b\partial\theta}
 \label{casimirsl2r}
\end{eqnarray}
It will serve us to complete higher-order polarization subalgebras.

As $SL(2,\mathbb R)$ is unimodular, left- and right-invariant integration
measures coincide. They have the form:
\begin{equation}
 d^Lg=\frac{da}{a^2}db d\theta=d^Rg.\label{measure}
\end{equation}

\subsection{Affine wavelets on $\mathbb R^+$ from discrete series of $SL(2,\mathbb R)$\label{sl2rRmas}}

We are seeking after $X=\mathbb R^+$ as the carrier space of the
representation of
$G$; this implies that $X^{L}_b$ and $X^L_\theta$ should be inside our polarization
subalgebra, but not $X^L_a$. Unfortunately, looking at the third
commutator in (\ref{sl2rcom}), it prevents us from a proper first-order
polarization. However, it is always possible to complete
$X^L_b$ with the Casimir operator (\ref{casimirsl2r}) and to consider the
second-order polarization subalgebra
${\cal P}_{\mathbb R^+}=<\hat{C}, X^{L}_b>$. The solution to the polarization
equations:
\begin{equation}
\left\{\begin{array}{l}
\hat{C}\Psi=q\Psi \\
X^{L}_b\Psi=\beta\Psi
\end{array} \right.
\end{equation}
has the form:
\begin{equation}
\Psi(a,b,\theta)=\,e^{\beta \,b }\psi(a,\theta)\label{sol1}
\end{equation}
where $\psi(a,\theta)$ satisfies the second order parabolic partial
differential equation:
\begin{equation}
\left[a^2\frac{\partial^2\ }{\partial a^2} +
a\beta\parcial{\theta}+(\beta^2 a^2-q)\right]\psi(a,\theta)=0
\label{Ec2grado}
\end{equation}

This is a parabolic partial differential equation similar to the Heat equation, which can be solved working out the value of $\parcial{\theta}\psi$, thus we can formally write the
solutions to this second-order partial differential equation as:
\begin{equation}
\psi(a,\theta)=e^{\theta\hat{\Theta}(a)}\varphi(a) \label{solHOP1}
\end{equation}
where
\begin{equation} \hat{\Theta}(a)\equiv -\frac{1}{\beta
a}\left[a^2\frac{\partial^2\ }{\partial a^2} +(\beta^2 a^2-q)\right] \label{solHOP2}
\end{equation}
is a second-order differential operator and $\varphi$ is an arbitrary
function of $a\in \mathbb R^+$. The action of the operators
${\cal X}$ (from right-invariant vector fields $X^R$) on solutions (\ref{sol1}), when
restricted to functions $\varphi$, turns out to be (after some algebra):
\begin{eqnarray}
{\cal X}_a\,\varphi(a)&\equiv&e^{-\beta b-
\theta\hat{\Theta}}iX^R_ae^{\beta b+ \theta\hat{\Theta}}\varphi(a)= i a\,\varphi'(a)
\nonumber\\ {\cal X}_b\,\varphi(a)&\equiv&e^{-\beta b-
\theta\hat{\Theta}}iX^R_be^{\beta b+ \theta\hat{\Theta}}\varphi(a)= i\beta\,
a\,\varphi(a) \label{solHOP3}\\
{\cal X}_\theta\,\varphi(a)&\equiv&e^{-\beta b-
\theta\hat{\Theta}}iX^R_\theta e^{\beta b+ \theta\hat{\Theta}}\varphi(a)=
i\hat{\Theta}(a)\,\varphi(a)
%\frac{-1}{\beta\, a} \left[ a^2\, \varphi''(a)-(\beta^2\, a^2+q)
%\varphi(a)\right]
\nonumber
\end{eqnarray}

As far as $\beta\neq 0$, the parameter $\beta$ does not play any role, and
can be eliminated defining $r=2\rho a$, with $\rho=i\beta$. With this change of variable, our
operators acquire the form:
\begin{equation}
{\cal X}_a= i r\,\frac{d}{d r},\;\;\;{\cal X}_b=
\frac{r}{2},\;\;\;{\cal X}_\theta= 2r\,\frac{d^2}{d r^2}
-\frac{r^2/2+2q}{r}. \label{op1}
\end{equation}

One can easily verify that the operators (\ref{op1}) define a
representation of the Lie algebra $sl(2,\mathbb R)$. Moreover,
the generators are hermitian  if the measure
$d\mu=dr/|r|$ is chosen and if $\rho\in \mathbb R$.
For $\rho>0$ the carrier space of the representation is $\mathbb{R}^{+}$, but for
$\rho<0$ is $\mathbb{R}^{-}$. This representation is irreducible and
it exponentiates to a unitary representation of the group $SL(2,\mathbb R)$ if the Casimir eigenvalue
verifies $q=k(k-1)$, with $k>0$ half-integer (the Bargmann index). For $k>\frac{1}{2}$ this
constitutes a representation of the discrete series of $SL(2,\mathbb R)$,
and therefore are square integrable representations.

For $\beta=0$, the expressions obtained have no meaning. In this case, the
solutions of Eq. (\ref{Ec2grado}) have a different form:
\begin{equation}
\psi(a,\theta)=a^{\alpha}\varphi(\theta)
\end{equation}

\noindent where $\alpha(\alpha-1)=q$. This case will be also recast later
in Sec. \ref{contser} as a solution of a different polarization leading to
the Continuous Series representation in the circumference. Thus, the
second order polarization ${\cal P}_{\mathbb R^+}$ encompasses both series
of representations.

We should note that this measure $d\mu=dr/|r|$ does not come directly from
(\ref{measure}) as $da/a^2=i_{X^L_b}i_{X^L_\theta}d^Lg$ does. The reason
is that ${\cal P}_{\mathbb R^+}$ is not a first-order, but a higher-order,
polarization.

We can construct a canonical basis for the Hilbert space $L^2_q(\mathbb R^+,dr/r)$
through eigenfunctions of $\X_\theta$ as follows:
\begin{equation}
-\frac{1}{2}\X_\theta\varphi_n^{k}(r)=(k+n)\varphi_n^{k}(r)\Rightarrow
\varphi_n^{k}(r)=e^{-r/2}r^k L^{2k-1}_n(r),
\end{equation}
where
%$q=k(k-1)$ is the Casimir eigenvalue ($k$ is the Bargmann index) and
$L^{m}_n(r)$ are generalized Laguerre polynomials fulfilling the standard equation:
\begin{equation} r\frac{d^2 L^m_n}{dr^2}+(m+1-r)\frac{d L^m_n}{dr} +n
L^m_n=0.
\end{equation}
Indeed, it can be easily proved that:
\begin{equation}
\langle\varphi^k_n|\varphi^k_m\rangle=\int_0^\infty
e^{-r}r^{2k-1}L^{2k-1}_n(r)L^{2k-1}_m(r)dr=\frac{(n+2k-1)!}{n!}\delta_{nm},
\,\,k\in \mathbb N/2.
\end{equation}
Thus, the set
\begin{equation}
{\cal B}_k=\left\{\langle r|k n\rangle=\phi_n^{k}(r)\equiv
\frac{1}{\sqrt{N^{k}_n}}\varphi_n^{k}(r) \right\},\,\,N^{k}_n=
\frac{(n+2k-1)!}{n!} \label{baseRmas}
\end{equation}
constitutes an orthonormal basis of $L^2_k(\mathbb R^+,dr/r)$.

Since the representations are square integrable (for half-integer $k>\frac{1}{2}$),
coherent states can be defined on the whole group, leading
to Klauder-Perelomov's coherent states (since these coherent states are invariant
modulo a phase under the Cartan subgroup, they can be defined on the
quotient $SL(2,\mathbb R)/U(1)$).

However, since we are interested in affine wavelets in $\mathbb R$, or $\mathbb
R^+$ in this case, we shall consider
the restriction of the whole representation
$U^L(g)$ of
$G$ to the affine subgroup, which consists of elements
$g=\sigma(a,b)=(a,b,\theta=0)$, and that can be obtained by exponentiation of the
infinitesimal generators $\X_a$ and
$\X_b$ (the factor $\frac{1}{2}$ appearing in $\X_b$ in (\ref{op1}) comes from the definition
$r=2\rho a$, which has been introduced to obtain the standard Laguerre polynomials, an
can be, of course,
eliminated). Doing so, we recover the usual irreducible representation of the
affine group on $L^2(\mathbb R^+,da/a)$, obtained in Sec. \ref{afinRmas}.
Note that representations of $G$ with different values of $k$ are
equivalent under the affine subgroup.

The construction of affine group wavelets $\gamma_{(a,b)}$ from coherent
states $\gamma_{(a,b,\theta=0)}$ of $G$ proceeds straightforwardly by taking
sections $\sigma(a,b)=(a,b,\theta=0)$.

\subsection{Affine wavelets on the right half complex plane and on $\mathbb R$}
\label{sl2rCmas}

For completeness, let us give the second-order polarization which provides
signals on the right half complex plane. It is: \be {\cal P}_{\mathbb
C^+}=<\hat{C}, X^{R}_\theta>.\ee We choose here right-(instead of
left-)invariant vector fields to simplify expressions. The solution to the
polarization conditions:
\begin{equation}
\left\{\begin{array}{l} \hat{C}\Psi=q\Psi \\ X^{R}_\theta\Psi=2 i k\Psi
\end{array} \right.\label{pol2}
\end{equation}
(the factor $2i$ in the second equation has been introduced for
convenience) has the form $\Psi(\theta,a,b)=e^{2\,i\,k \,\theta
}\psi(a,b)$, where
$\psi(a,b)$ satisfies the second order elliptic (Poisson-like) partial differential
equation:
\begin{equation}
\left[a^2\left(\frac{\partial^2}{\partial a^2} +\frac{\partial^2}{\partial
b^2} \right)+2ika\frac{\partial}{\partial
b}-q\right]\psi(a,b)=0.\label{schr}
\end{equation}
The general solution to this elliptic equation can be obtained by a
factorization method using complex variables. The solution is given in
terms of arbitrary analytic functions $\varphi$ of
$w=a+ib$ as $\psi(w,\bar{w})=(\Re(w))^k\varphi(w)$. The compatibility of the polarization conditions (\ref{pol2})
reproduces the relation $q=k(k-1)$.

The operators realizing the representation are given now by left-invariant
vector fields acting on wave functions defined on the whole
$SL(2,\mathbb{R})$ group. This representation, when restricted to
holomorphic functions $\varphi(w)$ on the half complex plane $\mathbb
C^+$, has the form:
\begin{eqnarray}
\X_a\,\varphi(w)&\equiv&a^{- k }\,e^{-2\,i\, k \,\theta }iX^L_aa^{ k
}\,e^{2\,i\, k \,\theta }\varphi(w)=i k \,\varphi( w) + i w \,\varphi'(
w)\nonumber
\\ \X_b\,\varphi( w)&\equiv&a^{- k }\,e^{-2\,i\, k \,\theta
}iX^L_ba^{ k }\,e^{2\,i\, k \,\theta }\varphi( w) =-\varphi'( w)
\\ \X_\theta\,\varphi( w)&\equiv&a^{- k }\,e^{-2\,i\, k
\,\theta }iX^L_\theta a^{ k }\,e^{2\,i\, k \,\theta }\varphi( w)= 2\,
 k \, w\,\varphi( w) -
  (  w^2-1) \,\varphi'( w)\nonumber
\end{eqnarray}
Again, this is an irreducible representation of $sl(2,\mathbb R)$, this
time on functions $\psi$ with support on $\mathbb C^+$. The generators are also
hermitian with the measure $d\mu=i_{X^L_\theta}d^Lg=a^{-2}dadb=\Re(w)^{-2}dw d\bar{w}$.
% which comes from (\ref{measure}) when we restrict ourselves
% to the affine subgroup.
The resulting Hilbert space $L^2_k(\mathbb C^+,d\mu)$, with scalar product
\begin{equation}
\langle \psi|\psi'\rangle=\int_{\mathbb
C^+}\bar{\varphi}(w)\varphi'(w)(\Re(w))^{2(k-1)}d\bar{w} d w,
\end{equation}
is isomorphic to the space $L^2_k(D,d\nu)$ of functions
$\psi(z,\bar{z})=(1-|z|^2)^k\varphi(z)$, [$\varphi(z)$ holomorphic], on the open unit
disk $D=\{z\in\mathbb C, |z|<1\}$ with integration measure
$d\nu=\frac{1}{(1-|z|^2)^2}d\bar{z}dz$. The transformation that maps the
half plane onto the disk is $z=\frac{w-1}{w+1}$. An orthonormal basis of
%$L^2_k(D,d\nu)$
$L^2_k(\mathbb C^+,d\mu)$ is given by:
%\begin{equation}
%{\cal B}_k=\left\{\phi_n^{k}\equiv \frac{1}{\sqrt{M^{k}_n}}(1-|z|^2)^k z^n
%\right\},\,\, M^{k}_n=\frac{n!(2k-2)!}{(2k+n-1)!}.
%\end{equation}
%
\bea
{\cal B}_k&=&\left\{\langle w| k n\rangle=\phi_n^{k}(w)\equiv
\frac{1}{\sqrt{M^{k}_n}} \Re(w)^k(1+w)^{-2k}
\left(\frac{w-1}{w+1}\right)^n \right\}\, \nn \\
 M^{k}_n&=&\frac{\pi
n!(2k-2)!}{2^{4k-2}(2k+n-1)!}. \label{baseCmas}
\eea
Denoting
$\tilde{\varphi}(w)=\langle w|\varphi\rangle$ and $\varphi(r)=\langle r|\varphi\rangle$, and inserting
the completeness relation $1=\int_{0}^\infty |r\rangle\langle r|dr/r$ we obtain a relation
\begin{equation}
\tilde{\varphi}(w)=\int_{0}^\infty \frac{dr}{r}\langle w| r\rangle \varphi(r)
\end{equation}
between functions in the half-plane representation,
$\tilde{\varphi}(w)\in L^2_k(\mathbb C^+,d\mu)$, and functions in the
``dilation'' representation,
$\varphi(r)\in L^2_k(\mathbb R^+,dr/r)$, with $\langle w| r\rangle$ the
kernel or ``polarization-changing operator'' (see Sec. \ref{PCO} and Sec. \ref{afinRmas}). An
explicit expression of this kernel can be calculated by inserting in
$\langle w| r\rangle$ the completeness relation for the eigenfunctions of
${\cal X}_\theta$,
$1=\sum_{n=0}^\infty |k n\rangle\langle k n|$, giving:
\begin{equation}
\langle w| r\rangle=\sum_{n=0}^\infty \phi_n^{k}(w)\bar{\phi}_n^{k}(r)=
\frac{1}{2\sqrt{\pi(2k-2)!}} \Re(w)^k r^ke^{-r\frac{w}{2}}
\end{equation}
where we have made use of the generating function of the generalized Laguerre
polynomials. Therefore, the ``polarization-changing operator" turns out to be:
\begin{equation}
\tilde{\varphi}(w)= \frac{1}{2\sqrt{\pi(2k-2)!}} \Re(w)^k \int_{0}^\infty
\frac{dr}{r} r^ke^{-r\frac{w}{2}} \varphi(r)
\end{equation}
This integral transformation is unitary, and is nothing other than the
Laplace transform. This is easily seen if we subtract the common weights
$\Re(w)^k$ and $r^{-(k-1)}$ from the wave functions:
\begin{equation}
(\Re(w)^{-k} \tilde{\varphi}(w))= [{\cal L}(r^{k-1}
\varphi(r))](w)=\frac{1}{2\sqrt{\pi(2k-2)!}} \int_{0}^\infty dr
e^{-r\frac{w}{2}} (r^{k-1} \varphi(r))
\end{equation}

In particular, the basis (\ref{baseRmas}) and (\ref{baseCmas}) are transformed
into each other under ${\cal L}$.

The second order elliptic equation (\ref{schr}) involves the variables $a$ and $b$, and we have solved
it in terms of the complex variable $w=a+ib$. In principle, we could solve it in terms of the variable
$a$, obtaining a realization in $\mathbb R^+$ equivalent to the one given in Sec. \ref{sl2rRmas}
(discrete series). We could also solve the equation in terms of $b$, obtaining a realization in $\mathbb R$
which would provide the discrete series of representations with support in $\mathbb R$. These
representations have a basis formed by the Relativistic Hermite Polynomials (since this representation can
be related to a model of a relativistic harmonic oscillator), which are directly related to the Gegenbauer
polynomials \cite{oscilata,oscipert}.

In this way, restricting to the affine subgroup, we could recover the affine wavelets in $\mathbb R$ given
in Sec. \ref{afinR}. However, we shall not pursue in this direction since the computations are involved
and the results have already been obtained in  Sec. \ref{afinR}.

%For those readers familiar with the language of quantum mechanics, we just
%shall comment that this holomorphic representation is the analogue of the
%Bargmann representation of a (relativistic) harmonic oscillator, where
%$\theta$ plays the role of ``time" parameter, $b$ is the ``position" and
%$a$ the ``momentum'' (see e.g. \cite{oscilata,oscipert}). Solving the ``relativistic Schr\"odinger''
%equation
%(\ref{schr}) in the ``position'' representation $(b)$, leads to a
%representation of the affine group on $L^2(\mathbb R)$ similar to the one
%given in Section \ref{afinR}, where a basis can be given in terms of
%Gegenbauer polynomials \cite{oscilata,oscipert}. In the same way, the
%solution in the ``momentum'' representation
%$(a)$ reproduces the representation of
%$SIM(1)$ on $L^2(\mathbb R^+)$ of the previous Section. Thus, the polarization ${\cal P}_{\mathbb
%C^+}$ encompasses in fact several carrier spaces. However, to avoid
%redundance, we shall not go into details. We just want to discuss in what
%follows the polarization changing operator between ${\cal P}_{\mathbb
%C^+}$ and ${\cal P}_{\mathbb R^+}$.

\subsection{Polarization $<\hat{C},X^L_a>$}
\label{polahyperbolic}

For the sake of completeness, we shall consider the polarization
${\cal P}^{HO}_{\mathbb S^1}=<\hat{C},X^L_a>$, although the solutions of this higher-order polarization
are the same as the ones for the first-order polarization that will be given in Sec. \ref{contser}.
The polarization equations are:
\begin{equation}
\left\{\begin{array}{l} \hat{C}\Psi=q\Psi \\ X^{L}_a\Psi=\alpha\Psi
\end{array} \right..
\end{equation}

The solution of the second equation is $\Psi(a,b,\theta)=a^\alpha
\phi(\tau,\theta)$, where
$\tau\equiv \frac{b}{a}$. The first equation then leads to:

\begin{equation}
\left[ ( 1 + \tau^2) \frac{\partial^{\,2}}{\partial \tau^2} +
 \frac{\partial^2}{\partial \theta\partial\tau}  +
         2\,\tau\,(1-\alpha) \,\frac{\partial}{\partial \tau}\right]\phi(\tau,\theta)=
(q - \alpha( \alpha-1) )\phi(\tau,\theta)\label{hyperbolic}
\end{equation}

This is an hyperbolic equation which, for $q = \alpha( \alpha-1)$ admits a very simple
factorization:
\begin{equation}
\left[(1 + \tau^2)\frac{\partial\,}{\partial \tau}+\frac{\partial\,}{\partial \theta}
-2(\alpha - 1)\tau  \right] [\frac{\partial\,}{\partial \tau}]\phi(\tau,\theta)=0
\end{equation}

Therefore its solutions are $\phi(\tau,\theta)=\phi(\theta)$. We shall not proceed further in the
details of this polarization since the solutions are the same as for the first-order polarization in Sec.
\ref{contser} (see below), even though we have in this case a second-order polarization. The reason lies
in the factorization of the second-order partial differential equation into a product of two first-order
differential operators. The solutions $\Psi(a,b,\theta)=a^\alpha \phi(\theta)$ really satisfy a couple of first-order
differential equations.

A different factorization of the equation (\ref{hyperbolic}) is possible, but with
$q=\alpha(\alpha+1)$. In this case the representation obtained has support not in $\theta$, but in the
variable $\theta-\arctan\tau$. The representation is, however, equivalent to the previous one and to the
representation obtained in Sec. \ref{contser}, leading to the continuous series
of representations with support on $\mathbb S^1$.

\subsection{Continuous Series in the circumference\label{contser}}
Now we are searching for $X=\mathbb S^1$ as the carrier space of the
representation of $G$; this implies that $X^{L}_b$ and $X^L_a$ have to be
inside our polarization subalgebra. Looking at the first commutator in
(\ref{sl2rcom}), we see that both vector fields already close a
first-order polarization subalgebra, which we shall denote by ${\cal
P}_{\mathbb S^1}=<X^{L}_a, X^{L}_b>$. The polarization conditions read:
\begin{equation}
\left\{\begin{array}{l} X^{L}_a\Psi=\alpha\Psi \\ X^{L}_b\Psi=\beta\Psi
\end{array} \right..
\end{equation}
Notice that, in particular, the commutation relations (\ref{sl2rcom})
force $\beta=0$, that is, the character of the polarization subalgebra is trivial on
the derived algebra, as explained in Sec. 3.1. The solution to these polarization
equations has the form:
\begin{equation}
\psi^\alpha(a,b,\theta)=W_\alpha(a,b,\theta)\,\gamma(\theta)=
a^{\alpha}\,\gamma(\theta)\,.
\end{equation}
Note that solutions $\psi$ do not depend on $b$. The operators realizing
the representation are given by right-invariant vector fields acting on
wave functions defined on the whole
$SL(2,\mathbb{R})$ group. Their restriction $\X=a^{-\alpha}iX^Ra^{\alpha}$ to
functions $\gamma$ supported on the circumference turns out to be:
\begin{equation}
{\cal X}_a=\frac{i}{2}\sin\,2\theta\,\frac{d}{d
\theta}+i\alpha\,\cos\,2\theta\,,\;\;\;{\cal X}_b=
\frac{i}{2}(\cos\,2\theta-1)\,\frac{d}{d \theta} -i\alpha\,\sin\,2\theta\,
,\;\;\;{\cal X}_\theta= i\frac{d}{d \theta}. \label{op2}
\end{equation}
%They are hermitian with respect to the projected measure
%$d\theta=i_{X^L_b}i_{X^L_a}d^Lg$ for $\alpha=\frac{1}{2} +is,\,\,s\in \mathbb{R}$.

%The representation (\ref{repre}) of $G$ on functions
%$\gamma(\theta)$ takes the form:
%%
%\be
%[{\calU}_\alpha(a',b',\theta')\gamma](\theta)\equiv
%a^{-\alpha}[U^L(a',b',\theta')(a^{\alpha}\gamma(\theta))]
%\ee

Following the general construction, the measure $d\mu$ can be computed as
$d\mu=i_{X^L_a}i_{X^L_b}d^Lg=\frac{d\theta}{a}$, which is again an ill-defined measure
on $\mathbb S^1$. To obtain an appropriate measure with the corresponding Radon-Nikodym
derivatives we must restrict to the values $\alpha=\frac{1}{2}+is\,,s\in\mathbb R$
(compare with the case of the affine group in Sec. \ref{afinR}),
and the resulting scalar product is given by:

\begin{equation}
\langle \Psi,\Psi'\rangle = \int_{-\pi}^\pi \frac{d\theta}{a} a^{\frac{1}{2}-is}
\bar{\gamma}(\theta)a^{\frac{1}{2}+is}\gamma'(\theta)=
\int_{-\pi}^\pi d\theta \bar{\gamma}(\theta)\gamma'(\theta)\,.
\end{equation}

These representations are however not irreducible, and decompose
into the direct sum of two irreducible representations, labelled
by the representations of the cyclic subgroup $\mathbb
Z_2=\{e,-e\}$ of $SL(2,\mathbb R)$ ($e$ denotes the identity
element). This means that, in order to achieve the complete
irreducibility, a non-connected polarization subgroup $P$ is
required. The representation associated with the trivial
representation of $\mathbb Z_2$ leads to the Continuous Series of
representations $C^0_q$, with $q=1/4+s^2$, of $SL(2,\mathbb R)$,
which are also representations of $SO(2,1)=SL(2,\mathbb
R)/{\mathbb Z}_2$ (see, for instance, \cite{Bargmann} and
\cite{Barut}). If we consider the non-trivial representation of
$\mathbb Z_2$, the representations obtained are the Continuous
Series $C^{1/2}_q$, which are representations of $SL(2,\mathbb R)$
but not of $SO(2,1)$\footnote{All these representations are
irreducible, except for $C^{1/2}_{1/4}$, with $s=0$, which
decomposes itself into two irreducible representations
\cite{Bargmann,Barut}.}.

Note that, since $\mathbb Z_2$ is central in $G$, imposing that
$\mathbb Z_2$ acts trivially in the finite right action [by the
polarization equation (\ref{pola-finita})], also means that it
acts trivially in the finite left action (\ref{regular}) and
therefore,  the transformation by $g=-e$ (a rotation by $\pi$)
acts trivially in the space of solutions of the representations
$C^0_q$; that is, a rotation by $\pi$ keeps the functions
unchanged, and therefore $\gamma(\theta)$ has periodicity $\pi$
instead of $2\pi$. This means that the true representation space
is $L^2((-\frac{\pi}{2},\frac{\pi}{2}),d\theta)$ instead of
$L^2((-\pi,\pi),d\theta)$ (see the comments below on the
definition of the dilation).

We shall restrict ourselves to the Continuous Series of
representations $C^0_q$, which are also representations of
$SO(2,1)$, and in particular to the representation $C^0_{1/4}$,
corresponding to $s=0$ and $\alpha=\frac{1}{2}$, denoting just the
operator ${\cal U}\equiv {\cal U}_{1/2}$. This representation (as
all the others in the Continuous Series) is not square integrable.
Thus, we proceed by restricting ourselves to the homogeneous space
$Q=G/N$, where $N$ is the subgroup of upper triangular matrices
with ones in the diagonal. Taking the Borel section
$\sigma(a,\theta)=(a,b=0,\theta)$, it is relatively easy to
compute the restricted finite group action ${\cal
U}(q)\gamma(\theta)\equiv
a^{-\frac{1}{2}}U^L(\sigma(q))a^\frac{1}{2}\gamma(\theta)$ of
elements $q=(a,\vartheta)\in G/N$ on functions $\gamma$ on
$\mathbb S^1$. In fact, using the group law (\ref{grouplaw}) to
compute $(a,0,\vartheta)^{-1}(1,0,\theta)$, after some algebra we
get:
\begin{equation}
\gamma_{\vartheta,a}(\theta)\equiv{\cal
U}_\alpha(a,\vartheta)\gamma(\theta)=
\lambda(1/a,\theta-\vartheta)^{\alpha}\gamma\left((\theta-\vartheta)_{1/a}\right),\label{reprecont}
\end{equation}
where
\begin{equation}
\theta_a\equiv\arctan(a\tan(\theta)),\;\;\lambda(a,\theta)\equiv\frac{a}{a^2+(1-a^2)\cos^2(\theta)},
\label{lambda}
\end{equation}
are the action $\theta\to\theta_a$ of dilations on elements
$\theta\in \mathbb S^1$ (see figure \ref{stereo} for a geometrical
interpretation of this transformation) and $\lambda$ is a
multiplier (\ref{multiplier}) which coincides with the
Radon-Nikodym derivative $\frac{d\theta_a}{d\theta}$, respectively. The action
of $G/N$ on $\mathbb S^1$ is defined modulo $\pi$,
%$\theta\in(-\frac{\pi}{2},\frac{\pi}{2})$,
as it was commented before. In fact, the action of the affine
subgroup $(a,b,0)$ is defined modulo $\pi$, as can be seen from
the group law (\ref{grouplaw}), the representation (\ref{op2}),
and Eqs. (\ref{reprecont}) and (\ref{lambda}), since $\theta$
appears always in the form of $\sin(2\theta)$ and $\cos(2\theta)$.
Therefore, we shall restrict ourselves in the sequel to this half
circumference. Note the difference between our projection from the
center of the (half-)circle and the case of the stereographic
projection from the south-pole to define a dilation on the sphere
in \cite{waveS2}. The reason for this difference is that we have
started with the group $SL(2,\mathbb R)$ instead of
$SO(2,1)=SL(2,\mathbb R)/{\mathbb Z}_2$, as would correspond to
obtaining wavelets on the circle in the paper \cite{nsphere},
where wavelets on the $(n-1)$-sphere were studied from the group
$SO_0(n,1)$ (note that the case of the circle is a singular case,
and this is why it does not fit in with the general scheme of
\cite{nsphere}). The purpose of using $SL(2,\mathbb R)$ instead of
$SO(2,1)$ is to study, with the same group and with the same
parametrization, wavelets on the circle and on the real line in a
unified manner, since the affine group can be obtained from
$SL(2,\mathbb R)$ by simply putting $\theta=0$ in the Iwasawa
decomposition of $SL(2,\mathbb R)$. The price we must pay for this
choice is that a quotient by $\mathbb Z_2$ must be done. This is
reflected in the fact that the minimal parabolic subgroup for
$SO(2,1)$ is connected meanwhile for $SL(2,\mathbb R)$ is
disconnected \cite{Barut}, containing a $\mathbb Z_2$ factor. In
fact, our polarization subgroup coincides with the minimal
parabolic subgroup in this case.

\begin{figure}[htb]
\begin{center}
\includegraphics[height=3.5cm,width=6cm]{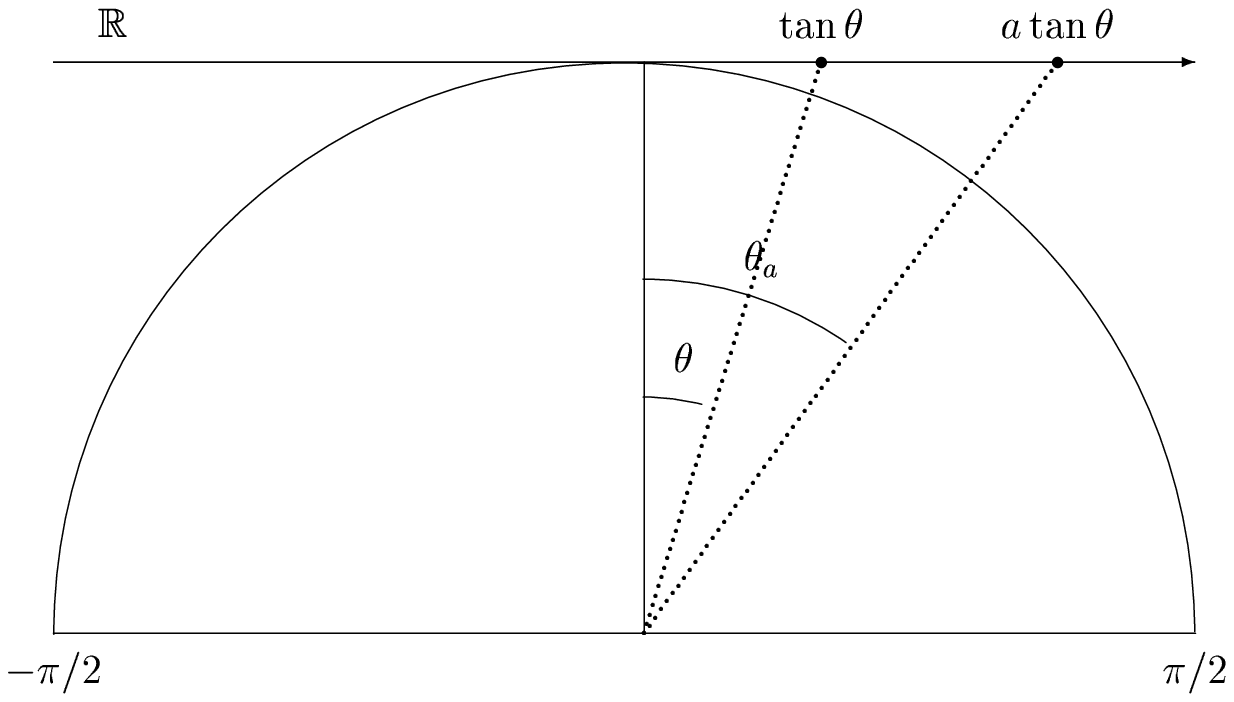}
\end{center}
\caption{Action of a dilation $\theta\to\theta_a$ on $(-\pim,\pim)$ lifted
from the tangent line $\mathbb R$ by inverse stereographic projection.}
\label{stereo}
\end{figure}

%The representation
%${\cal U}_\alpha$ corresponds to the continuous series of
%representations of $SL(2,\mathbb R)$ (see e.g. \cite{Barut}). It is well
%known (and it is not difficult to check) that the representation ${\cal
%U}_\alpha$ on $L^2((-\frac{\pi}{2},\frac{\pi}{2}),d\theta)$ is unitary
%just for values of $\alpha=\frac{1}{2}+is, \,s\in\mathbb R$. From now on,
%we shall restrict ourselves to the case $s=0$ and simply put ${\cal
%U}_{1/2}\equiv{\cal U}$ .

\subsubsection{Admissibility condition}
Let us investigate the general admissibility condition (\ref{qsquare}) for
our particular case. Here
\begin{equation}
d^Lq=i_{X^L_b}d^Lg=\frac{da}{a^2}d\theta
\end{equation} is the measure on
$Q=G/N$ projected from the measure (\ref{measure}) on the whole group. We
shall prove that:
\begin{thm}\label{basictheorem} The representation
${\cal U}$ given in (\ref{reprecont}) is square integrable
mod$(N,\sigma)$ and a non-zero vector
$\gamma\in L^2((-\frac{\pi}{2},\frac{\pi}{2}),d\theta)$  is admissible
mod$(N,\sigma)$, iff there exists a constant $c>0$ such that the quantity
\begin{equation}
\Lambda_n=\int_0^\infty\frac{da}{a^2}|\hat{\gamma}_a^n|^2<c,\,\,\forall
n\in\mathbb Z,\label{admistrong}
\end{equation}
where $\hat{\gamma}_a^n=\langle n|\gamma_a\rangle$ represents the Fourier
coefficient of $\gamma_a(\theta)=[{\cal U}(a,0)\gamma](\theta)$ in the
orthonormal basis
$\langle\theta|n\rangle\equiv\frac{1}{\sqrt{\pi}}e^{2in\theta}$ of
$L^2((-\frac{\pi}{2},\frac{\pi}{2}),d\theta)$.
\end{thm}
\ni\textbf{Proof.} Let us rewrite the general condition (\ref{qsquare}) in
our particular case. One can easily check that the transformation ${\cal
U}(\theta,a)$ can be split into two parts ${\cal U}(a,\theta)={\cal
U}(1,\theta){\cal U}(a,0)$. The effect of ${\cal U}(1,\theta)$ on the
Fourier coefficient of (\ref{reprecont}) is just a multiplicative phase
factor, that is:
\begin{equation}
\hat{\gamma}^n_{\vartheta,a}=\langle n|{\gamma}_{\vartheta,a}
\rangle=\frac{1}{\sqrt{\pi}}\int_{-\pim}^{\pim}
{\gamma}_{\vartheta,a}(\theta) e^{-2in\theta} d\theta=e^{-2in\vartheta}
\hat{\gamma}^n_{a}.
\end{equation}
Using the Parseval's identity
$1=\frac{1}{\pi}\sum_{n=-\infty}^\infty|n\rangle\langle n|$ for the Fourier series on
$(-\pim,\pim)$, the integrand of (\ref{qsquare}) can be written as:
\begin{equation}
|\langle {\cal
U}(\vartheta,a)\gamma|\psi\rangle|^2=\frac{1}{\pi}\sum_{m,n}
e^{2i(n-m)\vartheta}\bar{\hat{\gamma}}^n_a\hat{\gamma}^m_a\hat{\psi}^n\bar{\hat{\psi}}^m,
\end{equation}
and the admissibility condition (\ref{qsquare}) reads: \begin{equation}
\int_0^\infty\frac{da}{a^2}\int_{-\pim}^{\pim}d\vartheta |\langle
\gamma_{\vartheta,a}|\psi\rangle|^2=\int_0^\infty\frac{da}{a^2}\sum_{n=-\infty}^\infty
|\hat{\gamma}^n_a|^2|\hat{\psi}^n|^2=\sum_{n=-\infty}^\infty
\Lambda_n|\hat{\psi}^n|^2, \label{admicircle}
\end{equation}
where we have used orthogonality relations for trigonometric functions,
and used the definition (\ref{admistrong}) of $\Lambda_n$. Taking into account
that $\{|\hat{\psi}^n|^2\}\in \ell^1(\mathbb Z)$, since
$\sum_{n=-\infty}^\infty |\hat{\psi}^n|^2=||\psi||^2$, the
admissibility condition (\ref{qsquare}) adopts the following form:
\begin{equation}
 \sum_{n=-\infty}^\infty  |\hat{\psi}^n|^2\Lambda_n<\infty,\;\;\forall \{|\hat{\psi}^n|^2\}\in \ell^1(\mathbb Z),
\end{equation}
which converges absolutely iff $\{\Lambda_n\}\in \ell^\infty(\mathbb Z)$. That
is, $\gamma$ is admissible iff $\Lambda_n<c$, with $c$ independent of $n$, which
proves the theorem.$\blacksquare$

We shall postpone until the end of this Section \ref{contser} the proof of
the existence of functions $\gamma$ satisfying the condition
(\ref{admistrong}).

The strong condition (\ref{admistrong}) is necessary and sufficient for
the admissibility of $\gamma$, but it entails cumbersome evaluations of
Fourier coefficients. We shall give a more maneuverable condition,
although it will be only necessary. \begin{prop} A function
$\gamma\in L^2((-\pim,\pim),d\theta)$ is admissible only if it fulfils the condition:
\begin{equation}
\int_{-\pim}^{\pim}\frac{d\theta}{\cos\theta}\gamma(\theta)=0.\label{admiweak}
\end{equation}
\end{prop}
\ni\textbf{Proof.} Let us assume that $\gamma(\theta)=0$ if
$|\theta|>\tilde{\theta}<\pim$ (i.e. the support of $\gamma$ is bounded away from $\pm\pim$). Then,
$\gamma(\theta_{1/a})=0$ if $|\theta_{1/a}|>\tilde{\theta}\Rightarrow |\theta|>\tilde{\theta}_a$, where
$\theta_a$ defined in (\ref{lambda}) is the transformed angle under dilations. Thus, the Fourier
coefficient $\hat{\gamma}^n_a$ in (\ref{admistrong}) is calculated as:
\begin{equation} \hat{\gamma}^n_a=\langle
n|\gamma\rangle=\frac{1}{\sqrt{\pi}}\int_{-\tilde{\theta}_a}^{\tilde{\theta}_a}
\lambda(1/a,\theta)^{1/2}\gamma(\theta_{1/a})e^{-2in\theta}d\theta.
\end{equation}
Performing the change of variable $\vartheta=\theta_{1/a}$ and taking into
account that $\frac{d\theta_{1/a}}{d\theta}=\lambda(1/a,\theta)$ (the
Radon-Nikodym derivative) and that
$\lambda(1/a,\theta_a)=\lambda(a,\vartheta)^{-1}$, we have
\begin{equation} \hat{\gamma}^n_a=\langle
n|\gamma\rangle=\frac{1}{\sqrt{\pi}}\int_{-\tilde{\vartheta}}^{\tilde{\vartheta}}
\lambda(a,\vartheta)^{1/2}\gamma(\vartheta)e^{-2in\vartheta_a}d\vartheta.
\end{equation}
Let us split the integral (\ref{admistrong}) into three parts: short
($\epsilon\ll 1$), intermediate and large scales: \begin{equation}
\int_0^\infty=\int_0^\epsilon+\int_\epsilon^{1/\epsilon}+\int_{
1/\epsilon}^\infty. \label{splitint}
\end{equation}For
$a\ll 1$ we can approximate
$\lambda(a,\vartheta)^{1/2}\approx \frac{\sqrt{a}}{\cos\vartheta}$.
Therefore, the integral over small scales,
$a<\epsilon\ll 1$
\begin{equation}
\int_0^\epsilon \frac{da}{a^2}|\hat{\gamma}_a^n|^2\approx
\frac{1}{\pi}\int_0^\epsilon \frac{da}{a}
\left|\int_{-\tilde{\theta}}^{\tilde{\theta}}
\frac{d\theta}{\cos\theta}\gamma(\theta)\right|^2, \label{split3}
\end{equation}
converges only if the weak condition (\ref{admiweak}) holds (in fact, the
result must be independent of $\tilde{\theta}$). At intermediate scales,
the second integral in (\ref{splitint}) is finite because the integrand
$\hat{\gamma}^n_a$ is a bounded continuous function on
$(\epsilon,1/\epsilon)$; indeed,
${\cal U}(a,0)$ is a strongly continuous operator and the scalar product
is also continuous. It just remains to study the behavior at large scales
$a\gg 1$. Performing the change of variable $a\to 1/a$ and using the
behavior at short scales, we can write:
\begin{equation} \hat{\gamma}^n_a\approx \frac{1}{\sqrt{\pi}}
\int_{-\tilde{\theta}_a}^{\tilde{\theta}_a}
\frac{\sqrt{1/a}}{\cos\theta}\gamma(\theta_{1/a})e^{-2in\theta}d\theta,\;\;
a\gg 1.
\end{equation}
The only large scale divergence in the third part of (\ref{splitint}) will
never be reached because $\gamma(\theta)=0$ outside
$(-\tilde{\theta},\tilde{\theta})$ by assumption. Finally, if we drop the restriction on the support of
$\gamma$, the condition (\ref{admiweak}) is just necessary, which proves the
proposition.$\blacksquare$

%Yo sigo sin ver esto con claridad. \textbf{Mas
%bien me parece que la condición debe ser $\gamma(0)=0$ para que en el limite
%$\lim_{a\to\infty}\gamma(\theta_{1/a})=\gamma(0)=0$ la madre mate a
%$1/\cos\theta$, que diverge en $\pm\pim$, ¿no?...me parece que las madre ``perro" %DOG lo cumplen...}

The necessary condition (\ref{admiweak}) is the compact counterpart of the
usual zero-mean condition $\int_{-\infty}^\infty dx \gamma(x)=0$ for
affine wavelets on $\mathbb R$. We shall show in Sect. \ref{euclidlimit}
that they coincide in the Euclidean limit (large radius).

\subsubsection{Continuous frame condition}

Taking into account the result (\ref{admistrong}), the admissibility
condition (\ref{qsquare},\ref{admicircle}) can be written as:
\begin{equation}
\int_0^\infty\frac{da}{a^2}\int_{-\pim}^{\pim}d\vartheta |\langle
\gamma_{\vartheta,a}|\psi\rangle|^2\leq c||\psi||^2.\label{admicirc}
\end{equation}
This means that the family $\{\gamma_{\vartheta,a}, (\vartheta,a)\in
(-\pim,\pim)\times \mathbb R^+\}$ is a continuous family of coherent
states. Moreover:
\begin{prop}\label{continuousf} For any even admissible vector $\gamma$, the family
$\{\gamma_{\vartheta,a}, (\vartheta,a)\in
(-\pim,\pim)\times \mathbb R^+\}$ is a continuous frame; that is, there
exist constants $0<c_1\leq c_2$ such that
\begin{equation}
c_1||\psi||^2\leq \int_0^\infty\frac{da}{a^2}\int_{-\pim}^{\pim}d\vartheta
|\langle \gamma_{\vartheta,a}|\psi\rangle|^2\leq c_2||\psi||^2,\;\;\forall
\psi\in L^2((-\pim,\pim),d\theta) .\label{lowerupperb}
\end{equation}
\end{prop}
Before tackling the proof of this proposition, we introduce a previous
lemma, following the strategy of Ref. \cite{waveS2}.
\begin{lem} (1) The
correspondence $S:L^2((-\pim,\pim),d\theta)\rightarrow L^2(\mathbb R,dx)$
defined by: \begin{equation}
\Gamma(x)=[S\gamma](x)=\frac{1}{\sqrt{1+x^2}}\gamma(\arctan
x)\label{stereomap}
\end{equation}
is an isometry and an unitary map.

\ni (2) Let us denote $\tilde{D}_a={\cal U}(a,0): \gamma(\theta)\to
\lambda(1/a,\theta)^{1/2}\gamma(\theta_{1/a})$ the dilation
(\ref{reprecont}) on
$(-\pim,\pim)$ and $D_a: f(x)\to \frac{1}{\sqrt{a}}f(x/a)$ the usual dilation on $\mathbb{R}$. Then the
intertwining relation $S \tilde{D}_a=D_a S$ holds.
\end{lem}
\ni Both statements can be proved by a direct computation. We are now in
condition to prove the proposition \ref{continuousf}.

\ni \textbf{Proof of Proposition \ref{continuousf} } It remains only to
prove the lower bound. We claim that the quantity defined in
(\ref{admistrong}) is strictly positive $\Lambda_n>0,\,\forall n\in\mathbb{Z}$.
Assume on the contrary that $\Lambda_m=0$ for some $m$. This is possible only if
$\hat{\gamma}^m_a=0,\,\forall a\in\mathbb{R}^+$. Let us write this Fourier coefficient in a different
form using the previous Lemma:
\begin{equation} \hat{\gamma}^m_a=\langle
m|\tilde{D}_a\gamma\rangle_{L^2(-\pim,\pim)}=\langle S m|S \tilde{D}_a
\gamma\rangle_{L^2(\mathbb{R})}=2\sqrt{a}\int_0^\infty \frac{dx}{x}f(x)
g(a/x), \label{convo}
\end{equation}
with
\begin{equation}
f(x)=\frac{1}{\sqrt{1+x^2}}\cos(2m\arctan x),\;\;g(a/x)=
\frac{1}{\sqrt{1+(a/x)^2}}\gamma({\rm arccotan}(a/x)),
\end{equation}
where we have used that $\gamma$ is even. Notice that the integral
(\ref{convo}) has the form of a convolution in $L^2(\mathbb{R}^+,dx/x)$.
Looking at $\hat{\gamma}^m_a$ in this way, one knows that the convolution
is zero for all $a\in \mathbb{R}^+$ only if $f$ or $g$ are identically
zero, which implies $\gamma=0$. Therefore $\Lambda_n>0$ for every $n$. To
see that $\Lambda_n$ is also bounded from bellow and $\Lambda_n>c(\gamma)$
independent of $n$, we first show that the main contribution to the
integral (\ref{admistrong}) comes from the region
$a\sim 1/|m|\ll 1$ with $|m|>M\gg 1$ ($M$ fixed). Indeed, the Fourier
coefficient
\begin{equation}
\hat{\gamma}_a^m=\frac{1}{\sqrt{\pi}}\int_{-\pim}^{\pim}
\gamma_a(\theta)e^{-2in\theta}d\theta=
\frac{1}{\sqrt{\pi}}\int_{-\pim}^{\pim}
\lambda(a,\theta)^{1/2}\gamma(\theta)e^{-2in\theta_a}d\theta
\end{equation}
is nearly zero for $|m|>M\gg 1$ ($M$ fixed) and $a\gg \frac{1}{|m|}$.
Then, the main contribution to $\hat{\gamma}_a^m$ for $|m|>M\gg 1$ must come
from the range
$a\in (\frac{1}{|m|}-\frac{1}{k|m|},\frac{1}{|m|}+\frac{1}{k|m|})$, with $k>1$.
Using the approximations
$\lambda(a,\vartheta)^{1/2}\approx \frac{\sqrt{a}}{\cos\vartheta}$ and
$\theta_a\approx a\theta$ for small $a$, we can write $\hat{\gamma}_a^m\approx \sqrt{a}
\kappa(\gamma)$ with $\kappa(\gamma)=\frac{1}{\sqrt{\pi}}\int_{-\pim}^{\pim}
\frac{\gamma(\theta)}{\cos(\theta)}e^{-2i\theta}d\theta$ independent of
$m$. Thus, for large $m$, the asymptotic behavior of
\begin{equation}
\Lambda_m=\int_0^\infty\frac{da}{a^2}|\hat{\gamma}_a^m|^2\gtrsim
|\kappa(\gamma)|^2\int_{\frac{1}{|m|}-\frac{1}{k|m|}}^{\frac{1}{|m|}+\frac{1}{k|m|}}\frac{da}{a}=|k(\gamma)|^2\ln(\frac{k+1}{k-1})
\end{equation}
(for some finite $k>1$) gives a strictly positive quantity independent of
$m$, which proves that
$\Lambda_m$ is bounded from bellow. Notice that, in case $\kappa(\gamma)$ happened to be zero
for a particular vector $\gamma$, we could always
take $a\sim |l/m|$, with $|l|\ll |m|\gg 1$, so that the new lower bound
$\kappa_l(\gamma)\not=0$ for some $l$, otherwise $\gamma=0$.$\blacksquare$

We believe the requirement of $\gamma$ to be even in Proposition
\ref{continuousf} is too strong and perhaps it can be relaxed to a weaker
condition as to have a non-zero even part.

As in Ref. \cite{waveS2}, we shall conjecture here that, for general
admissible vectors, the frame $\{\gamma_{\vartheta,a}, (\vartheta,a)\in
(-\pim,\pim)\times \mathbb R^+\}$ is not tight. To get a tight frame
would require an equality in (\ref{lowerupperb}). Thus, we also believe
that the operator $A_\sigma$ in (\ref{pbiop}) has a spectrum in a
nontrivial interval, which should contract to a point in the Euclidean
limit (see next Section). Nevertheless, there must be particular vectors
$\gamma$ for which one can construct tight frames, but we shall not
discuss them here.

Now we are in condition to resume the question of the existence of
admissible circular vectors $\gamma$ by resorting to the well
known Euclidean case. First we note that:
\begin{lem}
The usual wavelet admissibility condition
$C_\Gamma=\int_{-\infty}^\infty|\hat{\Gamma}(k)|^2\frac{dk}{|k|}<\infty$
for the Fourier transform $\hat{\Gamma}(k)=\frac{1}{\sqrt{2\pi}}
\int_{-\infty}^\infty e^{-ikx}\Gamma(x)$ of functions $\Gamma\in
L^2(\mathbb{R})$ on the real line, is equivalent to the condition
\begin{equation} I_\Gamma=\int_{-\infty}^\infty db\int_{0}^\infty
\frac{da}{a^2}|\langle \Gamma_{b,a}|f\rangle|^2_{L^2(\mathbb{R})}
<\infty\label{admireal}
\end{equation}
for any $f\not= 0$ in $L^2(\mathbb{R})$, where
$\Gamma_{b,a}(x)=a^{-\um}\Gamma(\frac{x-b}{a})$ was defined in Eq.
(\ref{hijareal}).
\end{lem}
\ni\textbf{Proof.} See, for example, the reference
\cite{Gazeau}$\blacksquare$

\ni Next we can see that:
\begin{prop}\label{stereoproy} Any admissible wavelet on $L^2(\mathbb{R})$
yields an admissible wavelet on $L^2(-\pim,\pim)$ by inverse stereographic
projection (\ref{stereomap}).
\end{prop}
\ni \textbf{Proof.} We shall follow the strategy of Ref.
\cite{Bogdanova}. Let us split the general unitary operator ${\cal
U}(a,\vartheta)$ in (\ref{reprecont}) into translation and
dilation as ${\cal U}(a,\vartheta)={\cal U}(1,\vartheta){\cal
U}(a,0)$ and simply denote $R_\vartheta={\cal U}(1,\vartheta)$ and
$\tilde{D}_a= {\cal U}(a,0)$, as before. Then, the admissibility
condition (\ref{admicirc}) of a wavelet $\gamma\in
L^2(-\pim,\pim)$ can be written as:
\begin{equation}
I_\gamma=\int_{-\pim}^{\pim}d\vartheta\int_0^\infty\frac{da}{a^2}|\langle
R_{\vartheta}\tilde{D}_a\gamma|\psi\rangle|^2_{L^2(-\pim,\pim)}=\int_{-\pim}^{\pim}d\vartheta
I_\gamma(\vartheta) <\infty, \;\;\forall \psi \in L^2(-\pim,\pim).
\end{equation}
Using that the stereographic projection operator (\ref{stereomap}) is an
isometry and an unitary map, and that $R_\vartheta$ is unitary, we have
that:
\begin{equation}
I_\gamma(\vartheta)\equiv \int_0^\infty\frac{da}{a^2} |\langle
S\tilde{D}_a\gamma|SR_{\vartheta}^{-1}\psi\rangle|^2_{L^2(\mathbb{R})}=
\int_0^\infty\frac{da}{a^2} |\langle
{D}_aS\gamma|SR_{\vartheta}^{-1}\psi\rangle|^2_{L^2(\mathbb{R})},
\end{equation}
where we have used the intertwining relation $S \tilde{D}_a=D_a S$ in the
last equality. If $\Gamma(x)\equiv [S\gamma](x)$ is an admissible wavelet
on $L^2(\mathbb{R})$, the integral $I_\gamma(\vartheta)$ converges since
$[SR_{\vartheta}^{-1}\psi]\in L^2(\mathbb{R})$ for any $\psi \in
L^2(-\pim,\pim)$. Moreover, for any $\psi \in L^2(-\pim,\pim)$, the
integral $I_\gamma(\vartheta)$ is a continuous bounded function of
$\vartheta\in[-\pim,\pim]$, due to the compactness of the interval $[-\pim,\pim]$. Therefore,
$I_\gamma$ also converges, which means that $\gamma$ is an admissible
``circular'' wavelet. $\blacksquare$

Interesting cases of admissible vectors are the
``difference-of-Gaussians''
$\gamma_\alpha(\theta)=\gamma(\theta)-[\tilde{D}_\alpha\gamma](\theta)$,
with $\gamma(\theta)=e^{-\tan^2\theta}$, which one can see (at least
numerically) that verifies the necessary (weak) condition
(\ref{admiweak}).

Another natural question is the covariance of the CWT. Using the group law
(\ref{grouplaw}) it is easy to see that the CWT on the circle is covariant
under motions on $(-\pim,\pim)$, but not under dilations. Indeed, the
composition of two elements of the form $(\theta,a,b=0)$ gives an element
with $b\not=0$.

\subsection{The Euclidean limit\label{euclidlimit}}
One would expect the wavelet transform on the circle to behave locally (at
short scales or large values of the radius) like the usual (flat) wavelet
transform. Let us see that this is the case.

\subsubsection{Contracting  $SL(2,\mathbb{R})$ to $\mathbb{R}^2\rtimes
\mathbb{R}^+$}
A contraction ${\cal G}'=\mathbb{R}^2\rtimes
\mathbb{R}^+$ of ${\cal G}=sl(2,\mathbb{R})$ along the (identity connected component or the)
minimal parabolic
subgroup $P_0=AN=SIM(1)=N\rtimes \mathbb{R}^+$ can be constructed through the
one-parameter family of invertible linear mappings
$\pi_R:\mathbb{R}^3\to\mathbb{R}^3, R\in [1,\infty)$ defined by:
\begin{equation}
\pi_R(X_a)=X_a,\;\pi_R(X_b)=X_b,\; \pi_R(X_\theta)=R^{-1}X_\theta,
\end{equation}
such that the Lie bracket of ${\cal G}'$ is:
\begin{equation}
[X,Y]'=\lim_{R\to\infty}\pi_R^{-1}[\pi_RX,\pi_R Y],
\end{equation}
with $[\cdot,\cdot]$ the Lie bracket (\ref{sl2rcom}) of
${\cal G}$. The resulting ${\cal G}'$ commutators  are:
 \begin{equation}
[X_a,X_b]'=X_b,\;\;[X_a,X_\theta]'=-X_\theta,\;\; [X_b,X_\theta]'=0.
\end{equation}
The contraction process is lifted to the corresponding Lie groups
considering the mapping:
\begin{equation}\begin{array}{rcl}
\Pi_R: \mathbb{R}^2\rtimes \mathbb{R}^+& \rightarrow & SL(2,\mathbb{R})
\\
(x,y,a) & \mapsto & \Pi_R(x,y,a)=(\theta=\arctan\frac{x}{R},b=y,a).
\end{array} \label{contractionmap}
\end{equation}
Next we have to restrict the contraction map (\ref{contractionmap}) to the
homogeneous (parameter) spaces $G/N=Q$ and $G'/N=SIM(1)$ (notice that the
group structure of the parameter space and covariance are restored after
contraction); that is, we have to quotient out the nilpotent subgroup $N$,
which is preserved under contraction. To that aim, we introduce a section
$\sigma': G'/N \to G'$ defined by $\sigma'(x,a)=(x,0,a)$, which combined
with the canonical projection of the Iwasawa bundle $p: G\to Q=G/N$, leads
to the natural restricted contraction map:
\begin{equation}\begin{array}{rcl}
\Pi_R^{\sigma'}: \mathbb{R}\rtimes \mathbb{R}^+_*&\rightarrow
&SL(2,\mathbb{R})/N
\\ q'=(b,a) &\mapsto &  q=\Pi_R^{\sigma'}(q')=p(\Pi_R(\sigma'(q')))=(\arctan\frac{b}{R},a).
\end{array}
\end{equation}

\subsubsection{The Euclidean limit of the CWT on the circle}
The Euclidean limit is formulated as a contraction at the level of group
representations. We shall see (the procedure parallels that of Ref.
\cite{waveS2}) that the continuous series representation of $G$ contract
to the usual wavelet representation of the affine group $SIM(1)$ in the
following sense:

\begin{defn}
Let $G'$ be a contraction of $G$, defined by the contraction map
$\Pi_R:G'\to G$, and let ${\cal U}'$ be a representation of $G'$ in a Hilbert space ${\cal H}'$.
Let $\{{\cal U}_R\}, R\in [1,\infty)$ be a one-parameter family of
representations of $G$ on a Hilbert space ${\cal H}_R$, and $I_R:{\cal
H}_R\to {\cal D}_R$ a linear injective map from ${\cal H}_R$ onto a dense
subspace ${\cal D}_R\subset {\cal H}'$. Then we shall say that ${\cal U}'$ is a
contraction of the family
$\{{\cal U}_R\}$ if there exists a dense subspace ${\cal D}'\subset{\cal H}'$  such that,
for all $f\in{\cal D}'$ and $g'\in G'$, one has:
\begin{itemize}
\item For every $R$ large enough, $f\in{\cal D}_R$ and
${\cal U}_R(\Pi_R(g'))I_R^{-1}f\in I_R^{-1}{\cal D}_R$ .
\item $\lim_{R\to\infty}||I_R{\cal U}_R(\Pi_R(g')I_R^{-1}f-{\cal U}'(g')f||_{{\cal H}'}=0,\;\;\forall
g'\in G'$.
\end{itemize}
\end{defn}

In our case, ${\cal H}_R=L^2[(-\pim R,\pim R),Rd\theta]\approx {\cal
H}=L^2[(-\pim ,\pim ),d\theta]$ is the Hilbert space of square integrable
functions on the half-circumference of radius
$R$ and ${\cal H}'= L^2(\mathbb{R},dx)$. It is easy to check that
\begin{eqnarray}
I_R: &{\cal H}_R\rightarrow& {\cal H}'\nonumber\\ \gamma &\to&
[I_R\gamma](x)=\frac{1}{\sqrt{1+(x/R)^2}}\gamma(\arctan(x/R))
\end{eqnarray}
is an isometry, i.e.:
$||I_R\gamma||_{{\cal H}'}=||\gamma||_{{\cal H}_R}$. The
inverse map
\[[I_R^{-1}f](\theta)=\frac{1}{\cos\theta}f(R\tan\theta)\]
is also an isometry. For each $R$, we choose
${\cal D}_R={\cal D}'=C_0(\mathbb{R})$, the space of continuous functions of compact
support. Then, we have to prove that:
\begin{thm}
The representation $[{\cal U}'(b,a)f](x)=\frac{1}{\sqrt{a}}f(\frac{x-b}{a})$ of
the affine group $SIM(1)$ is a contraction of the one-parameter family
${\cal U}_R$ of representations of $SL(2,\mathbb{R})$
on ${\cal H}_R$, given by (\ref{reprecont}) realized on ${\cal H}_R$, as
$R\to\infty$. That is:
\begin{equation}
\lim_{R\to\infty}||I_R{\cal U}_R(\Pi_R^{\sigma'}(b,a))I_R^{-1}f-{\cal U}'(b,a)f||_{{\cal
H}'}=0,\;\;\forall (b,a)\in \mathbb{R}\rtimes \mathbb{R}^+
\end{equation}
\end{thm}
\noindent\textbf{Proof.} By direct computation
\begin{equation}
\lim_{R\to\infty}|I_R{\cal U}_R(\Pi_R^{\sigma'}(b,a))I_R^{-1}f-{\cal U}'(b,a)f|=0,\;\;
\forall (b,a)\in \mathbb{R}\rtimes \mathbb{R}^+
\end{equation}
which means pointwise convergence.

Now consider $f\in {\cal D}'$. Since $f$ is continuous and of compact support, it is bounded.
Since $I_R$, ${\cal U}'$ and ${\cal U}_R$ are unitary, the expression
\begin{equation}
|I_R{\cal U}_R(\Pi_R^{\sigma'}(b,a))I_R^{-1}f-{\cal U}'(b,a)f|<|I_R{\cal U}_R(\Pi_R^{\sigma'}(b,a))
I_R^{-1}f|+|{\cal U}'(b,a)f|
\end{equation}
is uniformly bounded and the left hand side converges pointwise to zero as $R\rightarrow \infty$.
Then Lebesgue's dominated convergence theorem states that it tends to zero in the strong sense.
$\blacksquare$

\section*{Conclusions and outlook}
We have exposed a general approach to obtain the CWT on a manifold
$X$ associated to a (first or higher-order) polarization subalgebra ${\cal
P}_X$ of a symmetry group $G$ of ``affine transformations". As an
application, we have derived the CWT on $\mathbb R$ and $\mathbb S^1$
entirely from the group $SL(2,\mathbb R)$. Theorem \ref{basictheorem}
yields the basic ingredient for writing a genuine CWT (\ref{cwt}) on
$\mathbb S^1$. We have provided admissibility conditions and proved that, for an
admissible even vector $\gamma$, the family $\{\gamma_{\vartheta,a}\}$ is
a continuous frame. The proposed CWT on the circle has the expected
Euclidean limit; that is, it behaves locally like the usual (flat) CWT. We
have given a precise mathematical meaning to these notions using the
technique of group contractions.

In this article we mainly focused on ``the continuous approach'',
based on the theory of coherent states of quantum physics
(formulated in terms of group representation theory). It remains
to study ``the discrete approach", which roots in the
Littlewood-Paley analysis, and yields fast algorithms for
computing the wavelet transform numerically. An intermediate
approach which paves the way between the continuous and the
discrete cases is based on the representations of some finite
groups \cite{Flornes,Holschneider-finite,discrete-dilation}. We
believe a thorough group-theoretic treatment of (finite) periodic
sampled signals is possible, where the classification of filters
amounts to a group-theoretic cohomology problem. However, the
discussion of this interesting subject deserves a separate study.

\end{document}